\documentclass[12pt]{article}
\usepackage{aaspp4,astrobib,epsf}
\usepackage{times,mathptm}

\def\etal{{\it et al.}}
\def\simlt{\hbox{ \rlap{\raise 0.425ex\hbox{$<$}}\lower 0.65ex\hbox{$\sim$} }}
\def\simgt{\hbox{ \rlap{\raise 0.425ex\hbox{$>$}}\lower 0.65ex\hbox{$\sim$} }}
\def\lta{\mathrel{\spose{\lower 3pt\hbox{$\sim$}}\raise 2.0pt\hbox{$<$}}}
\def\gta{\mathrel{\spose{\lower 3pt\hbox{$\sim$}}\raise 2.0pt\hbox{$>$}}}

\def\psd{deg$^{-2}$}

\lefthead{Nikolaev \& Weinberg}
\righthead{LMC populations from 2MASS}

\begin{document}

\title{Stellar Populations in the Large Magellanic Cloud from 2MASS}
\author{Sergei Nikolaev \& Martin D. Weinberg}
\affil{University of Massachusetts, Amherst MA 01002}
\date{\today}

\begin{abstract}
 
  We present a morphological analysis of the feature-rich 2MASS LMC
  color-magnitude diagram, identifying Galactic and LMC populations 
  and estimating the density of LMC populations alone.  We also 
  present the projected spatial distributions of various stellar 
  populations.  The conditions that prevailed when
  2MASS observed the LMC provided $10\sigma$ limiting sensitivity
  for $J\simlt16.3$, $H\simlt15.3$, $K_s\simlt14.7$.
  Major populations are identified based on matching 
  morphological features of the color-magnitude diagram with 
  expected positions of known
  populations, isochrone fits, and analysis of the projected spatial
  distributions.  

  2MASS has detected a significant 
  population of asymptotic giant branch (AGB) stars ($\simgt 10^4$ 
  sources) and obscured AGB stars ($\simgt 2 \times 10^3$ sources).  
  The LMC populations along the first-ascent red giant branch (RGB) 
  and AGB are quantified.  Comparison of the giant luminosity
  functions in the bar and the outer regions of the LMC shows 
  that both luminosity functions appear consistent with each other.  
  The luminosity function in central (bar) field has well-defined 
  drop-off near $K_s=12.3^m$ corresponding to the location of RGB 
  tip; the same feature is seen in the luminosity function of the 
  entire LMC field.

  Isochrone fits for the corresponding giant branches reveal no
  significant differences in metallicities and ages between central and
  outer regions of the LMC.  This may be evidence for strong dynamical
  evolution in the last several gigayears.  In particular, the observed 
  LMC giant branch is well-fit by published tracks in the CIT/CTIO system 
  with a distance modulus of $\mu=18.5\pm0.1$, reddening $E_{B-V}=0.15-
  0.20$, metallicity $Z=0.004^{+0.002}_{-0.001}$ and age $3-13$ Gyr.  
  Analysis of deep 2MASS engineering data with six times the standard 
  exposure produces similar estimates.  
\end{abstract}


\section{Introduction} \label{sec:introduction}

Large and homogeneous data sets of near-infrared photometry for the 
entire LMC, a by-product of large-scale infrared sky surveys such as 
2MASS \cite{skr97} and DENIS \cite{epc97}, have become available to 
astronomical community only recently.  Cioni \etal{} (2000) introduced 
the DENIS Point Source Catalog towards the Magellanic Clouds.  Here, 
we present the LMC data from The Two Micron All Sky Survey (2MASS).  

The 2MASS has observed the entirety of the
Large Magellanic Cloud and much of these data are included in
the recent second incremental data release.  Empirically, the 
photometry has signal-to-noise (SNR) ratio 10 at J, H, K$_s$ magnitudes of 
16.3, 15.3, 14.7, respectively, slightly better than the nominal survey 
limit.  At these limits, we can observe all of the thermally-pulsing 
asymptotic giant branch (AGB) populations and part way down the red 
giant branch (RGB).  The red clump,
representing helium burning giants, lies $\sim 2$ mag below the 
sensitivity limit of these data.
The extinction in near-infrared (NIR) is small throughout the LMC and 
negligible on average everywhere but the inner degree of arc.  Together 
with the high quality of 2MASS photometry ($\sigma_m \sim 0.03^m$), 
overall zero-point stability (better than $0.01^m$) and with reliable 
identification of LMC populations, the survey is ideal for studies of 
spatial structure of the LMC or its evolution (see the companion paper, 
Weinberg \& Nikolaev 2000; hereafter WN).  

We describe data selection,
cross-correlation with a few well-known populations,
and comparison of the 2MASS color system in \S\ref{sec:data}.  We 
present a morphological analysis the color-magnitude diagram in 
\S\ref{sec:cmd}.  In particular, we make identifications of the Galactic 
and LMC populations corresponding to all features in the color-magnitude 
diagram, correlate these with spatial distribution, and estimate the 
density of LMC populations alone.  The LMC giant branch is well 
determined and we separately identify the AGB, first-ascent red giant 
branch tip (TRGB), and the carbon star sequence.  The luminosity 
function of RGB and AGB populations is derived (\S\ref{sec:lf}) and 
compared with the galactic giant-branch luminosity function.  We explore 
the feasibility of determining the spatial dependence of metallicity 
using giant-branch morphology.  Finally, \S\ref{sec:summary} summarizes 
our results and discusses the implications and opportunity for future
study.

\section{Data} \label{sec:data}

Our LMC field is $\approx 250$ sq. degrees and covers the range from 
$4^h00^m$ to $6^h56^m$ in right ascension and from $-77^\circ$ to 
$-61^\circ$ in declination (coordinates in J2000.0).  The initial sample
of $7,092,894$ sources is drawn from the Working Survey Data Base and 
includes possible artifacts, such as filter glints and diffraction spikes 
from nearby bright stars, source confusion, and detection upper-limits.  The 
known contaminants and flux statistics are well-characterized and 
identified during processing \cite{cut99}.  Eliminating artifacts and 
requiring detections at all bands with $\sigma_m \le 0.11^m$ (SNR $\ge 10$) leaves 
$1,246,304$ stars.  Unlike the released catalog, these data contain 
multiple apparitions sources because of scan overlaps.  We identify the 
multiple entries based on 1)~spatial proximity ($|\Delta r|\le 2''$), 
and 2)~matching $K_s$-photometry ($|\Delta r|\le 5\sigma$).  Note that
this procedure differs from that of the 2MASS catalog release (see
Cutri \etal{} 1999).  Our final 
sample contains $823,037$ sources.  

\begin{figure}[h]
\epsfysize=12.0cm
\centerline{\epsfbox{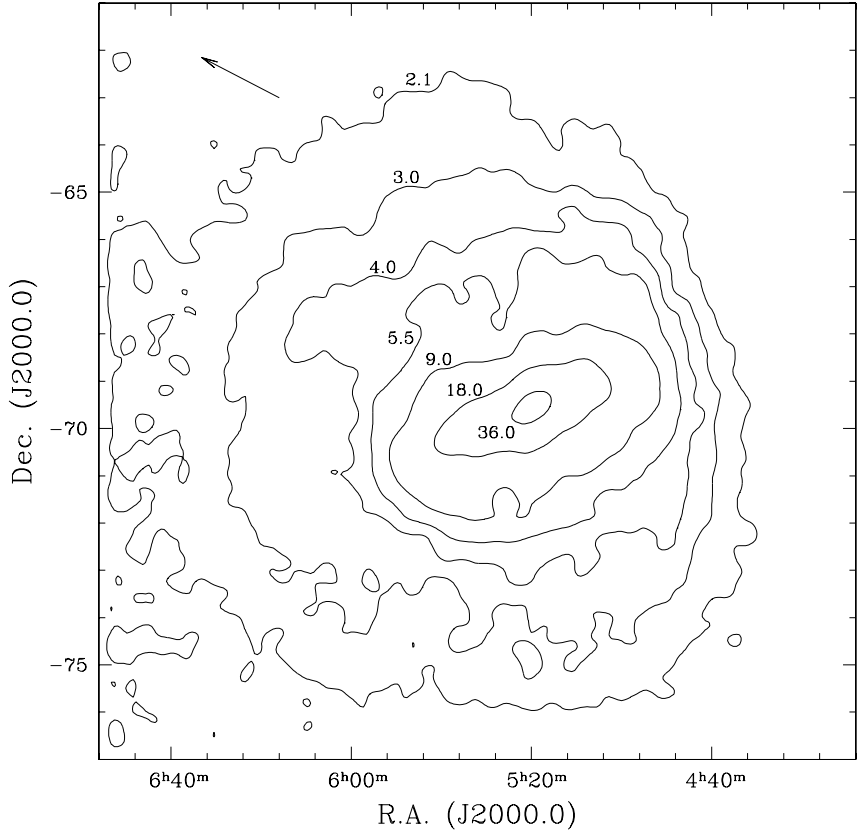}}
\caption{Distribution of 2MASS sources in the LMC field.  Contour
  levels are labeled in units of $10^3$ \psd.  Arrow points in the
  direction to Galactic center. \label{fig:sky}}
\end{figure}

The spatial distribution of these 
stars is shown in Figure~\ref{fig:sky}.  The figure shows major 
structural components of the LMC, the bar and disk, immersed in the 
field of Galactic foreground.  The gradient of the foreground sources 
(the direction to the Galactic center is indicated by an arrow) distorts 
the isopleths of the outer LMC disk.
The source density near the optical 
center of the bar (at $\alpha=5^h24^m$, $\delta_{J2000}=-69^\circ44'$) 
exceeds $3.6 \times 10^4$ \psd{}.  The expected mean separation among
sources at this density, $\zeta = 1/\sqrt{\rho} \sim 20''$, is much 
greater than 2MASS resolution and therefore confusion should not be 
significant (see also Wood 1994).  As a separate check, we have examined 
the source counts as the function of magnitude in dense central fields 
and in more sparse fields in the outer LMC.  The distribution of source 
counts as a function of magnitude have similar shapes in both dense and 
sparse regions, consistent with low confusion.

We have cross-correlated our sources 
with the database of long-period variables (LPV) from Hughes \& 
Wood~(1990).  Their sample includes 376 large-amplitude Mira-like 
variables and 224 smaller amlitude semi-regular variables.  We find 370 
($98.4\%$) and 224 2MASS counterparts, respectively\footnote{The search 
radius extends to $5''$ from the listed positions of sources.}.
Three of the `missing' large-amplitude variables are
present in the raw 2MASS data, but are degraded by artifacts (two 
diffraction spikes and one blend).  Of the remaining three, two are 
matched by stars of the appropriate magnitudes within $7''$ radius, and 
only one does not have any match to within $20''$ radius.  All 134 
Wolf-Rayet stars in the LMC \cite{bre99} have been observed by 2MASS 
\cite{vdk99}.

\subsection{$K_s$ vs. $K$ Photometry} \label{sec:Kshort}

The 2MASS photometric system is similar to CIT/CTIO system
\cite{eli82}, except that it uses $K_s$ band ($2.00-2.32\, \mu\rm{m}$)
rather than $K$ band. The $K_s$ (`K-short') bandpass is described by
Persson \etal{} (1998).  It was designed to reduce the ground-based
thermal background.  The transmissivity curve for the filter is given
in Persson \etal{}, who also compared $K_{CIT}$ with $K_s$ photometry
for a set of solar-type stars and red standards (see their Tables 2
and 3, respectively).  Based on their data, the difference $K-K_s$
shows no significant systematic trend in the color range $0 < J-K <
3$.  The strongest trends follow from the presence of CO-band
absorption, which affects the $K$ filter more than the $K_s$
filter.  The absolute value of the difference $|K-K_s|$ is less than
$0.05^m$ and we will assume $K_s = K_{CIT}$ in comparing CIT/CTIO
system-based stellar sequences with 2MASS data.

\subsection{Interstellar Reddening} \label{sec:reddening}

One of the advantages of 2MASS as compared to optical surveys is low 
interstellar reddening, since extinction at $2 \mu$m is approximately 10 
times smaller than in $V$.  The values of interstellar reddening 
$E_{B-V}$ found in the literature\footnote{foreground plus LMC internal 
reddening} fall in the range between $0.08$ \cite{mat87} and 
$0.20$ \cite{har97}.  The distribution of reddening from Harris \etal{} 
(1997) has non-Gaussian tail to high values. Greve \etal{} (1990) have 
reported values as high as $E_{B-V} = 1.1$, found from their 
investigation of dust in emission nebulae in the LMC.  Bessell (1991) 
has summarized reddening determinations from photometry, stellar 
polarization and HI column densities to derive foreground and intrinsic 
mean reddening in the Clouds.  He obtained typical LMC internal 
reddening of $0.06$ (with substantial variations), and foreground 
reddening in the range $0.04- 0.09$.  Galactic foreground reddening can 
be surprisingly large in the outer regions of the LMC: Walker (1990) 
reported $E_{B-V}=0.18\pm0.02$ at NGC 1841, about $15^\circ$ from the 
optical center of the LMC.  On the other hand, in the cluster GLC 0435-59 
(Reticulum), $11^\circ$ from the LMC center, the reddening is only 
$E_{B-V}=0.03$ \cite{wal92}.  Zaritsky (1999) has indicated that 
reddening for F and G stars in the LMC is much less than that for OB stars
and derived the average $<E_{B-V}>=0.03$ for late-type stars in the disk.

In the present study, the data are not dereddened.  Rather, each diagram 
shows the direction and magnitude of the reddening vector for a specified
value of $E_{B-V}$.  The reddening vector is based on relations from
Koornneef (1982):
\[
A_K = 0.189 E_{B-V}; \quad E_{J-K} = 0.651 E_{B-V},
\]
for $R = A_V/E_{B-V} = 3.1$.  Information about the reddening 
may be obtained directly from 2MASS data from the analysis 
of the color-color diagram (see below). 
 
\section{Analysis of the Color-Magnitude Diagram} \label{sec:cmd}

\begin{figure}[t!]
\mbox{
\mbox{\epsfysize=8.3cm\epsfbox{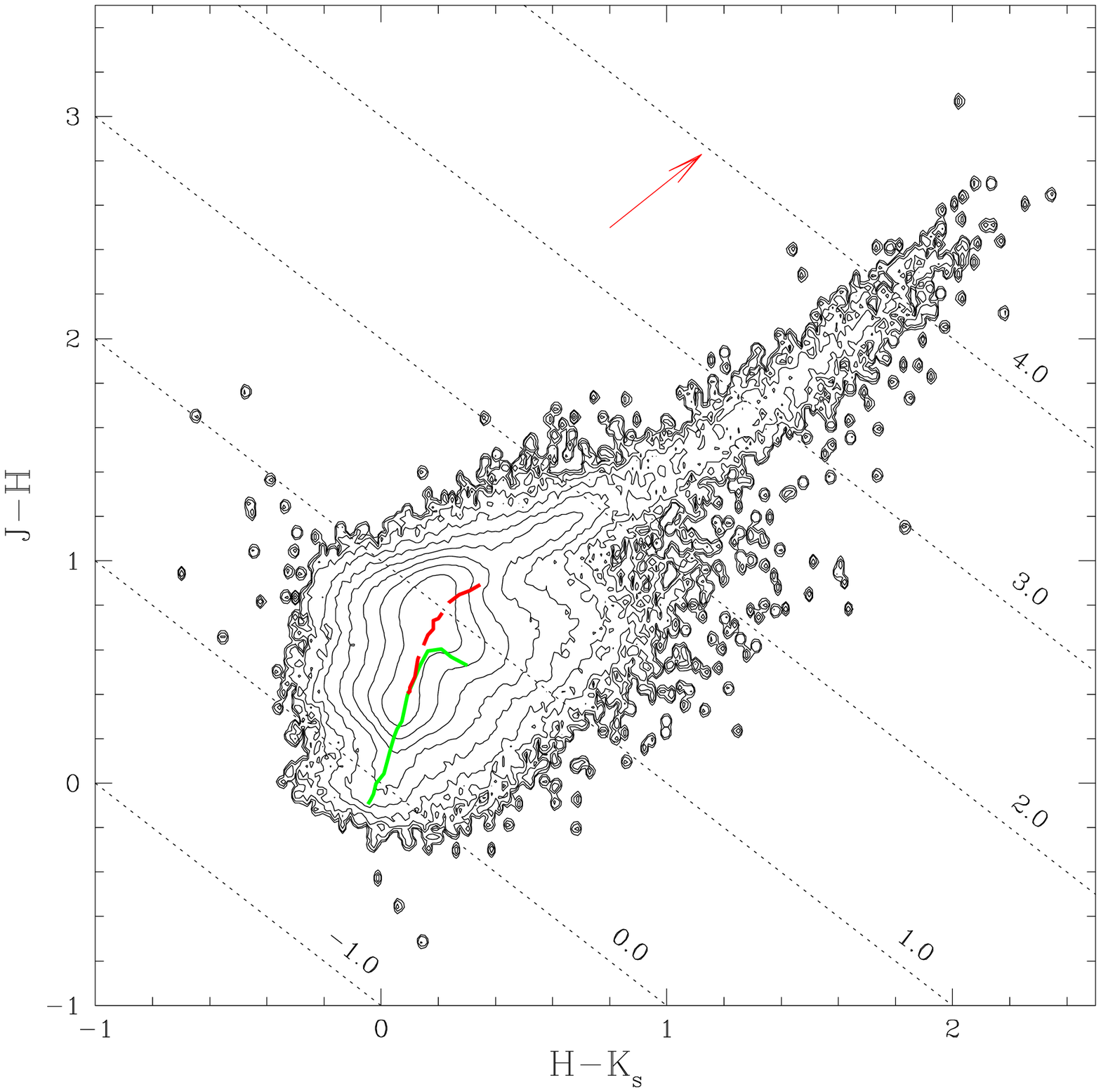}}
\mbox{\epsfysize=8.3cm\epsfbox{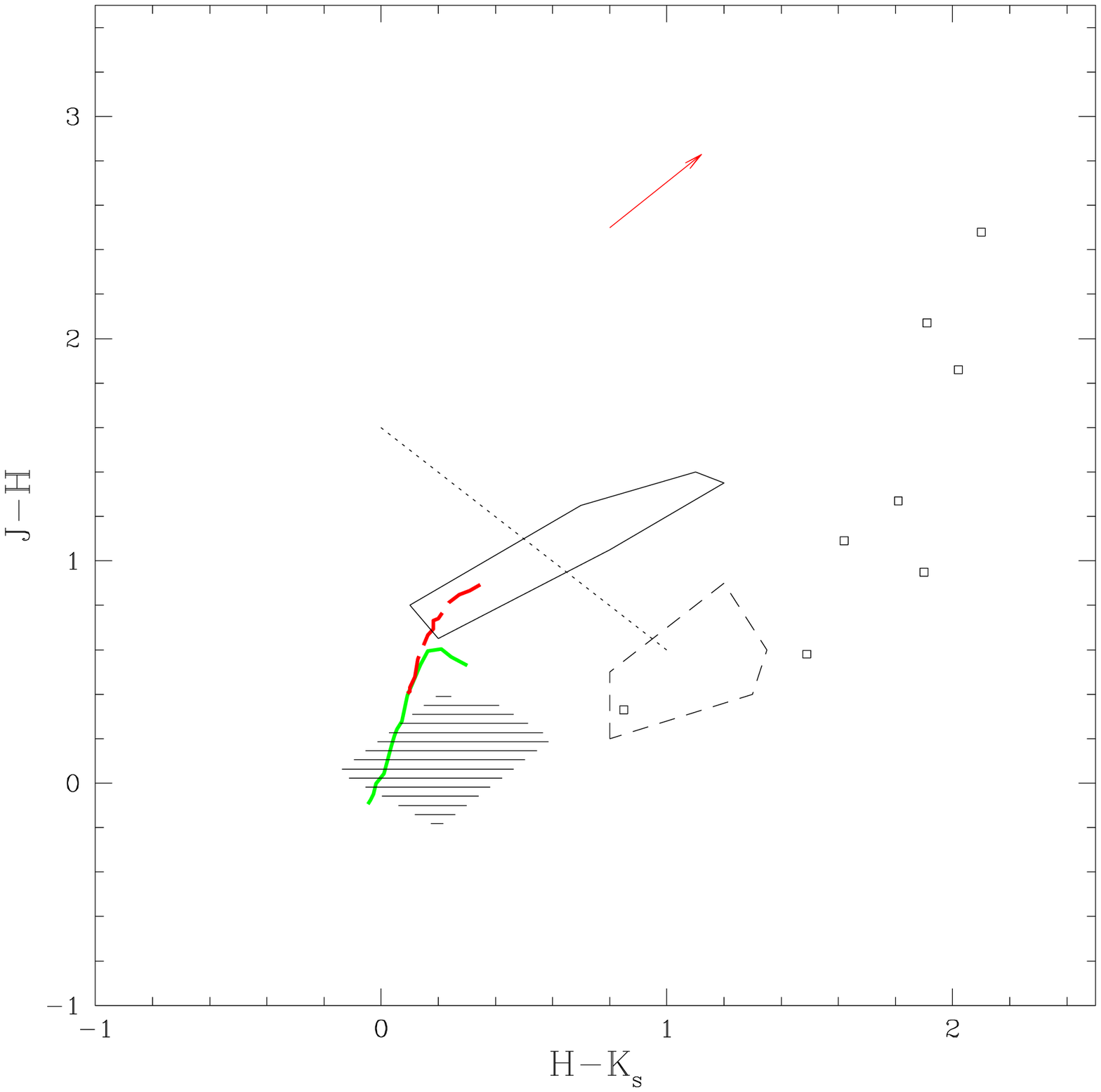}}
}
\caption{{\em Left panel}: Color-color diagram of the LMC field.  
  Contour levels are logarithmic, from 2 to 6.5, spaced by 0.5.  
  Diagonal lines are lines of constant $J-K_s$ values (marked).  
  Color sequences of dwarfs (solid line) and giants (dashed line) 
  from W92 are shown.  Reddening vector for $E_{B-V} = 1.0$ 
  (indicated by arrow) is based on relations from Koornneef~(1982), 
  assuming $R = 3.1$. {\em Right panel}: Color-color diagram showing 
  approximate positions of some LMC populations.  Shaded area 
  corresponds to Wolf-Rayet stars (Breysacher 1999); region outlined 
  by dashed lines encompasses known LMC B[e] stars (Gummersbach~\etal{} 
  1995), open squares show individual observations of four LMC 
  protostars (from Westerlund 1997).  W92 fiducial colors of dwarfs 
  (thick solid line) and giants (thick dashed line) are indicated.  
  Region occupied by carbon stars in the sample of Costa \& Frogel 
  (1996) is shown with solid lines.  Dotted line corresponds to 
  $J-K_s=1.6$.  Reddening vector is drawn for $E_{B-V}=1.0$.
\label{fig:ccd}}
\end{figure}

Figure~\ref{fig:ccd} shows the color-color ($J-H$ vs. $H-K_s$) diagram
of the LMC for $823,037$ 2MASS sources selected in \S\ref{sec:data}.  
The diagram shows relatively few distinct features.  The most prominent 
among them the extended `arm' of the thermally-pulsing AGB stars 
(TP-AGB) in the upper right 
corner.  Typical colors of LMC M giants are in the range 
$0.2\simlt H-K_s\simlt0.3$, $J-H\simgt 0.8$ \cite{fro90}.  Most stars 
on the extended arm, with colors redder than $J-K_s=1.6$ are
carbon-rich stars \cite{hug90}.  Of those with the extremely red 
colors, $J-K_s>2.0$, many probably possess dusty circumstellar envelopes.  
The sample contains $\sim 2000$ such sources. Their locus is consistent with 
the track following the reddening vector for $H-K_s > 1.0^m$.
Figure~\ref{fig:ccd} also shows fiducial color tracks 
for both giants and dwarfs from Wainscoat \etal{} (1992; hereafter W92).  
The two sequences overlap near $H-K_s=0.15$, $J-H=0.5$, since NIR 
colors of late G --- early M type dwarfs are the same as colors of late 
F --- early K type giants.  

Because of the small number of features and the general compactness
of the color-color diagram, its usefulness in discriminating the
major populations is limited, especially in the
overlap region, $0.5<J-K_s<0.8$.  However, some LMC populations which 
occupy distinct regions in the diagram can still be identified based 
of their location in the color-color plot (see Figure~\ref{fig:ccd}).  
In particular, the color-color diagram is quite useful in 
identifying candidate sources with large infrared excess, such as 
young protostars, cocoon stars, or obscured AGB carbon stars.

The color-color diagram may be used to determine the reddening distribution.
The giant population forms a tight branch in 
near-infrared colors.  The reddening direction nearly coincides with 
$J-K_s$ and therefore the distribution in $J-K$ of a sample from 
narrow color interval in $J-H$ along the giant branch provides a 
sensitive diagnostic for reddening by dust.  We considered a sample
of sources in the range $0.78<J-H<0.85$.  The resulting shift of the 
peak is $\Delta(J-K_s)<0.03$ ($\Delta(J-K_s)<0.06$) outside (inside) of
the central region, suggesting only minor reddening on
average on scales larger than 0.1 square degrees.

\begin{figure}[t!]
\mbox{
\mbox{\epsfysize=8.3cm\epsfbox{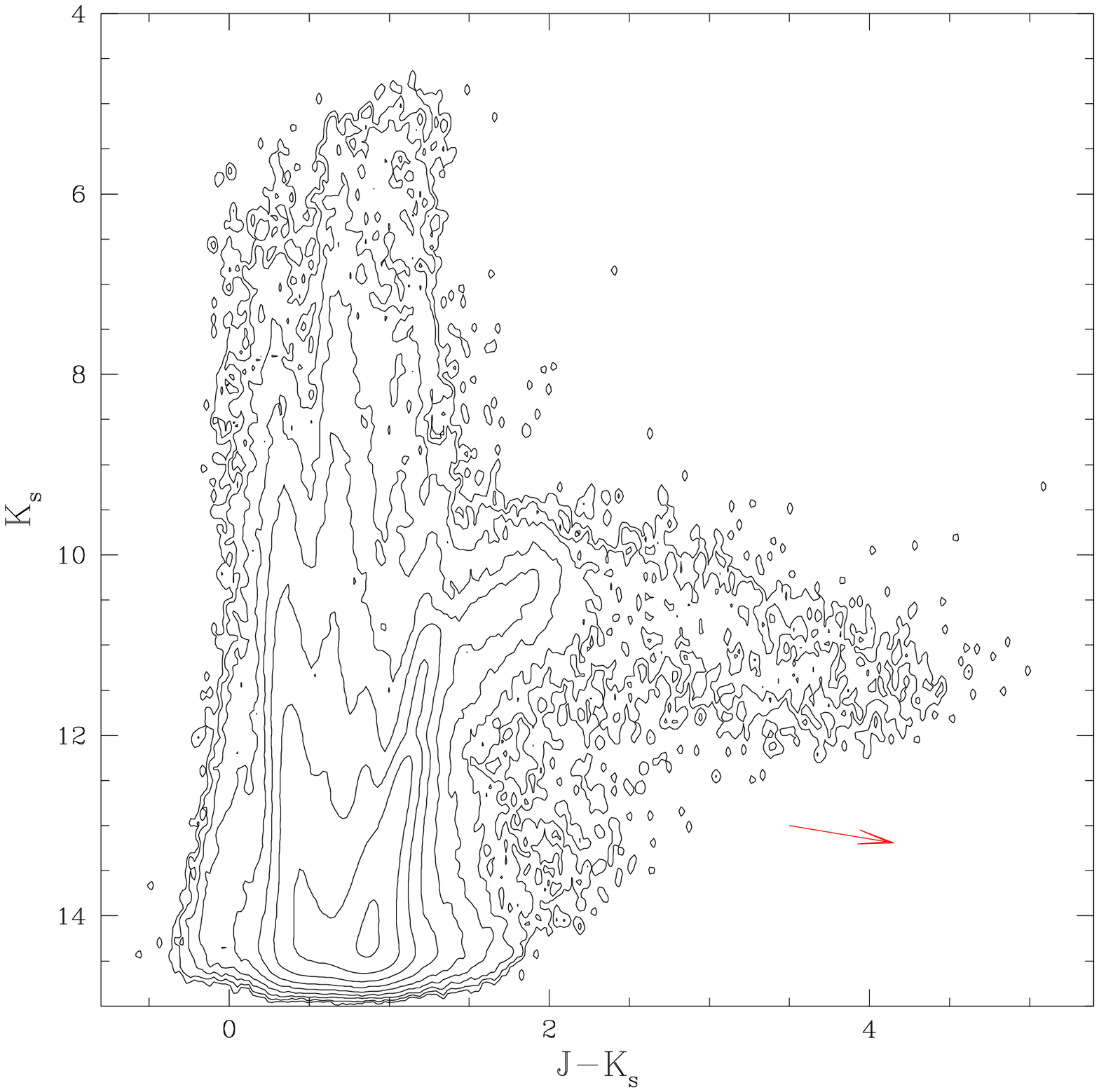}}
\mbox{\epsfysize=8.3cm\epsfbox{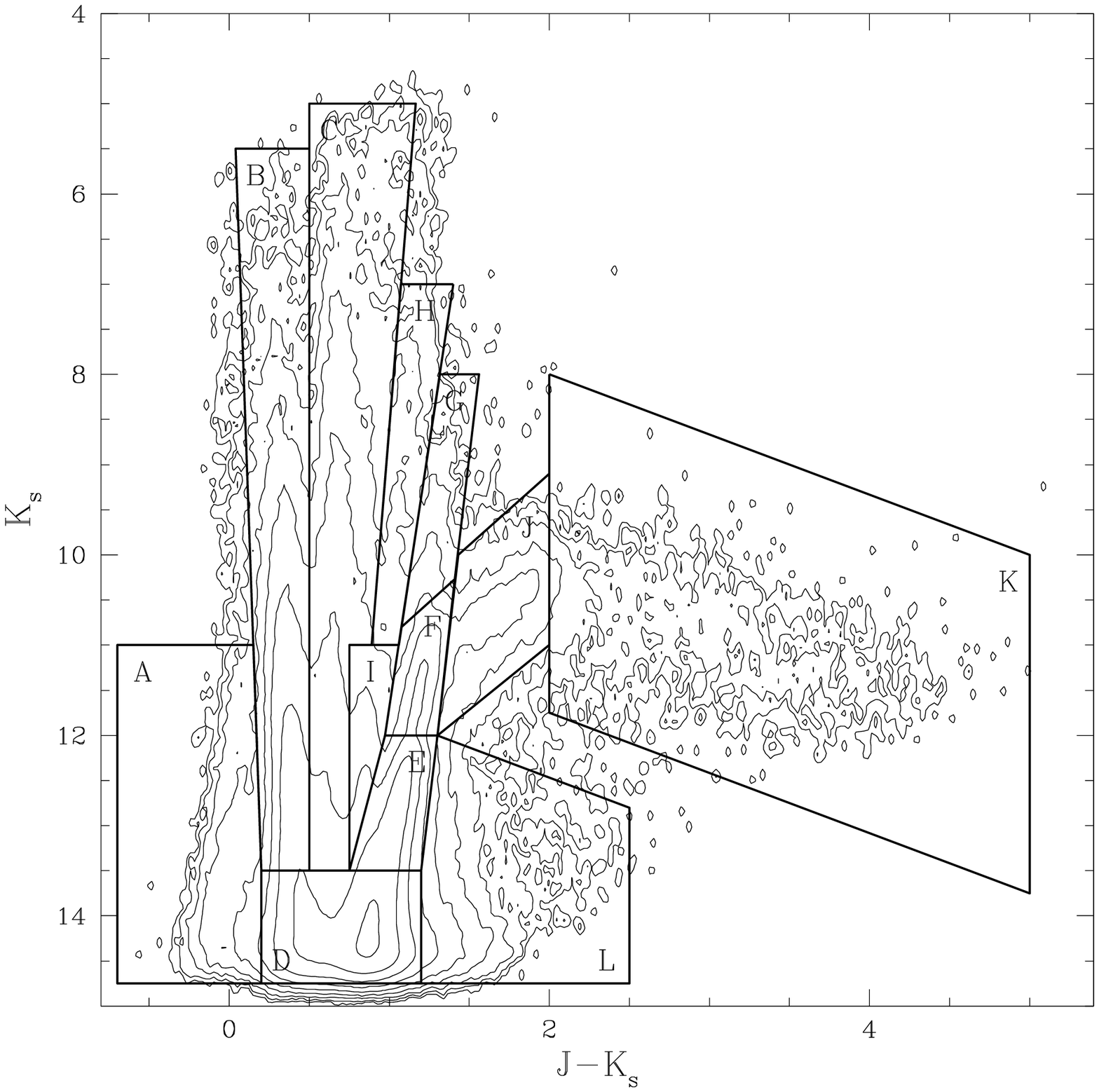}}
}
\caption{{\em Left panel}: Color-magnitude diagram of the LMC field.
  The density levels are logarithmic, from 2 to 6, spaced by 0.5.  
  The reddening vector corresponds to $E_{B-V} = 1.0$. {\em Right
  panel}: The same diagram with highlighted 12 regions discussed in
  text.  The regions correspond to major features of the CMD.
\label{fig:cmdregs}}
\end{figure}

\begin{figure*}
\epsfysize=20.0cm
\centerline{\epsfbox{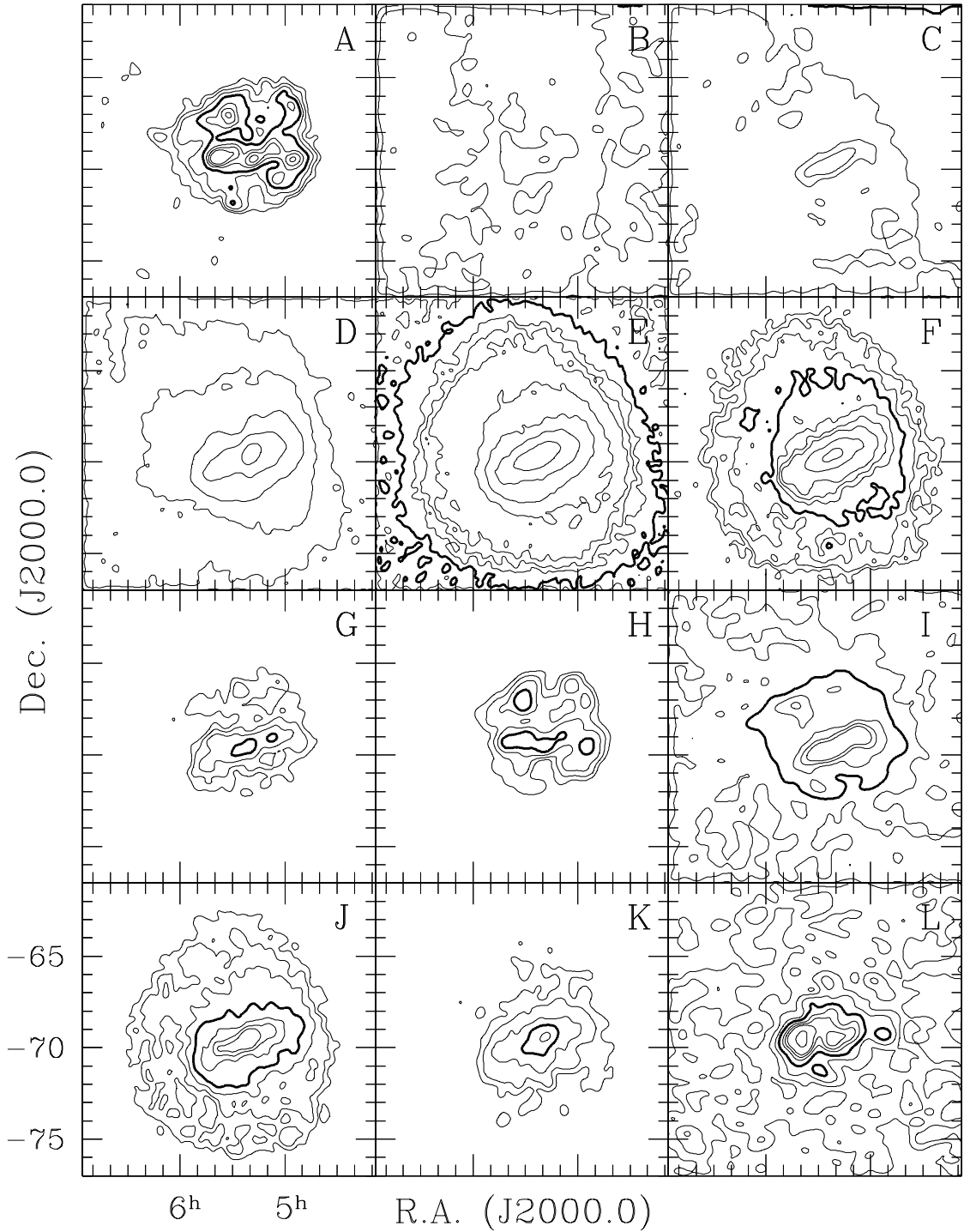}}
\caption{Spatial density distributions of sources in CMD regions.  
  Letters correspond to the regions introduced in 
  Figure~\ref{fig:cmdregs}.  The distributions are kernel smoothed 
  with a Gaussian kernel.  The same sequence of contour levels is 
  used in each panel (see text).  Contour level of 120 \psd{} is 
  highlighted. 
\label{fig:skyregs}}
\end{figure*}

The color-magnitude diagram (CMD)
presented in Figure~\ref{fig:cmdregs} reveals a wide variety of 
details.  Our goal is reliable identification criteria for LMC stellar 
populations based on their positions in the diagram.  The CMD is
hand-shaped with vertically-stretched `fingers' (e.g., at $J-K_s$ 
colors of 0.4, 0.6, 1.1) due to varying distance modulus for both 
Galactic and Magellanic sources.  We have identified 12 regions shown 
in Figure~\ref{fig:cmdregs} that highlight distinct features of the 
CMD.  The regions are marked A through L and enclose $99.7\%$ of the 
$823,037$ sources in the field.  To identify stellar populations in 
each region, we use a combination of several techniques.  The 
Galactic foreground contribution is modeled by a synthetic CMD based 
on the tabulated near-infrared model of W92.  The LMC populations are 
identified based on the infrared photometry of known populations found 
in the literature.  In addition, we do a preliminary isochrone 
analysis, where we match the features of the CMD with Girardi \etal{} 
(2000) isochrones to derive the ages of populations and draw 
evolutionary connections among the CMD regions.  The details of our 
population matching procedure are given in \S\ref{sec:identify}.  In 
addition, we use the spatial density distributions of sources in each 
region to better discriminate local and LMC populations.  The spatial 
distributions for each region are shown in Figure~\ref{fig:skyregs}.
In each panel of the figure, we plot source density contour levels of 15, 
30, 60, 120, 240, 350, 480, 960, 1920, 3840, 7680, and 15000 \psd{}.  The
contour levels are selected both to show the underlying density profiles 
with maximum details and facilitate the comparison among different 
population densities.  However, due to strong variations in relative 
density in the CMD, not all contour levels in this sequence are
displayed: in panels E and I the lowest density contour corresponds to 60 
\psd{}, in panels B and C the contours start from 240 \psd{}, and in very 
dense Region~D the lowest contour level is 960 \psd{}.  

\subsection{Identifying Stellar Populations} \label{sec:identify}

The initial analysis of the CMD regions has two parts: 1)~use of the 
spatial density distribution to estimate the location (Galaxy or LMC); and 
2)~use of the theoretical colors/isochrones to derive the properties of 
the population, such as age, approximate spectral class and distance
modulus.  Here we describe the procedure of identifying the populations,
before examining each region of CMD in detail.

\begin{figure}[h]
\epsfysize=9.0cm
\centerline{\epsfbox{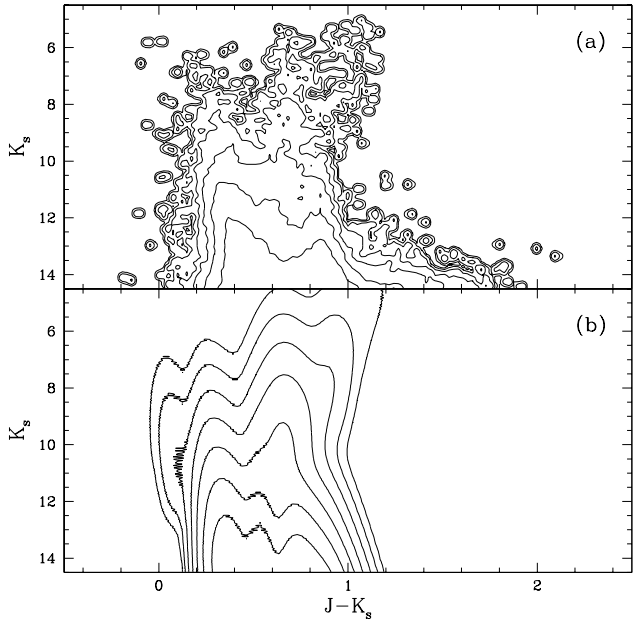}}
\caption{Comparison of (a) observed CMD of a Galactic field and (b) the 
synthetic Galactic CMD from W92 model, as explained in text.  Densities 
in both diagrams are normalized to unity.  The contour levels are 
logarithmic, from $-0.5$ to $-3.5$, spaced by 0.5.  Similar plots for each 
stellar class and for each structural component of the Milky Way allows 
unambiguous determination of populations responsible for observed CMD 
features. \label{fig:cmds}}
\end{figure}

The CMD of the LMC field (Fig.~\ref{fig:cmdregs}) contains both Galactic 
and LMC populations.  To quantify Galactic foreground, we use near-infrared model 
of W92, based on 8-25 micron point source counts.  Galactic model of W92 
has five structural components: exponential disk, bulge, stellar halo, 
spiral arms and molecular ring.  The main contribution to the Galactic 
source density toward the LMC ($l=280.5$; $b=-32.9^\circ$) 
is the exponential disk.  The source density due to other Galactic 
structural components combined does not exceed $0.0005\%$ in any region 
of the CMD.  The luminosity function in W92 model is represented by a sum of
stellar classes, allowing independent estimate of the contribution of 
each class to the CMD.  Each class of source is assumed to have a 
Gaussian distribution,
\[
N (M) \propto \exp \left[ -\frac{(M-M_\lambda)^2}{2\sigma^2} \right].
\]
To model the Galaxy, we use the first 33 classes from Table~2 of W92
(Galactic dwarfs, giants and supergiants).  The remainder (AGBs, planetary
nebulae, etc.) are expected to give only a small contribution to source 
density and thus do not affect the CMD.  In the Galactic model, we use 
reddening parameters from Rieke \& Lebofsky (1985).  Dust is assumed to 
follow a double exponential distribution, with the radial scale length of 
the disk and the scale height of 100 pc.  We reduced the magnitude dispersion 
$\sigma$ of each stellar class by ten to accurately represent the CMD.  
Reducing $\sigma$ results in a spiky 
differential luminosity function but this does not affect our application 
and the cumulative luminosity function remains well-approximated.  Given 
the granularity of the model, the agreement between original W92 luminosity 
function and ours is acceptable.  Our synthetic `foreground' CMD is shown 
in Figure~\ref{fig:cmds}, along with the observed CMD of a Galactic field
\footnote{To define our Galactic field, we combined three small fields, each 
$\approx 0.6$ sq. deg.  These fields were selected at the boundary of our 
LMC field where contamination from LMC source is minimal. At the time, other
contiguous fields were not available.}.  The agreement 
between the model and observed CMD is good, with few easily
explainable discrepancies.  For example, the extension of the CMD at 
$J-K_s\simgt1$, $K_s\approx13-14$ is due to the population of field 
galaxies \cite{jar98}.

Conclusions about the LMC populations in the CMD are made based on
isochrone fitting or on empirical matching of populations found in the
literature to features of the CMD.  We use theoretical isochrones from
Girardi \etal{} (2000).  These isochrones supersede Bertelli's set
\cite{ber94}, and use updated opacities and equations of state.  The
isochrones follow the evolution of low- and intermediate-mass stars 
($0.15 M_\odot < M < 7 M_\odot$) from the main sequence up to the tip 
of the RGB or the start of the thermally-pulsing AGB.  
Distinct LMC populations are 
identified by matching morphological features of the CMD with colors 
of known populations from the literature.  In particular, we use 
Cepheid colors from Madore \& Freedman (1991), early M supergiant 
color sequence from Elias \etal{} (1985), and data on long-period 
variables from Hughes \& Wood (1990).  This matching is purely 
qualitative and is only used as a supplement.  The
fiducial colors of Galactic giants and supergiants from W92 model are
unsuitable for LMC, since LMC has a lower metallicity compared to the
Milky Way.  Moved to the LMC distance, $\mu=18.5$, the W92
giant branch provides a poor fit to the observed RGB (see
Figure~\ref{fig:match_g} below).

%%%%%%%%%%%%%%%%%%%%%%%%%%%%%%%%%%%%%%%%%%%%%%%%%%%%%%%%%%%%%%%%%%%%%%%%%%%%%%%
%                       REGION___A
%%%%%%%%%%%%%%%%%%%%%%%%%%%%%%%%%%%%%%%%%%%%%%%%%%%%%%%%%%%%%%%%%%%%%%%%%%%%%%%
\subsection{Region A: Blue Supergiants, O Dwarfs} \label{sec:A}

These blue-colored sources are readily identified as early type Population~I 
stars in the LMC.  This group of stars is the evidence of recent 
($<30$~Myr) star formation.  Plotting the theoretical evolutionary 
tracks in the CMD (Figure~\ref{fig:match_d}) confirms that the region is 
populated by blue supergiants and brightest dwarfs (ZAMS).  Only the hottest 
and most massive dwarfs of types O3---O6 can be seen in the LMC at $\mu=18.5$.
All other main sequence (MS) populations are too faint and fall below the 
limit at SNR $=10$ imposed for this work.  The supergiant population in 
the region are core 
helium-burning stars with masses $4 \simlt M \simlt 9 M_\odot$.  These stars 
spend most of their post-MS lifetimes as blue or red supergiants \cite{mae89}
looping between Regions~A and H of the CMD (cf. Figure~\ref{fig:match_d}b).  
Region~A encompasses stars which are at the blue tips of their blue loops.
While crossing Regions~B and C, these stars enter the instability strip and 
become Cepheids (see \S\S~\ref{sec:B}, \ref{sec:C}).  

The spatial distribution of these objects also clearly indicates an
LMC population.  The distribution is rather clumpy with several richest OB
associations outlining the location of spiral arms and brightest and largest 
HII regions (e.g., 30~Dor).  The density concentrations in 
Fig.~\ref{fig:skyregs}A are consistent with the well-known superassociations 
and Shapley's Constellations \cite{mar76,vdb81}.  These youngest populations 
do not trace the bar of the LMC, in agreement with de~Vaucouleurs~\& Freeman 
(1973).  Quantitative analysis of the distribution \cite{wei00} puts the 
centroid of the population at $\alpha = 5^h23^m$, $\delta = -68^\circ48'$, 
about $1^\circ$ north of the optical center of the bar.

The Galactic population of early dwarfs is readily seen in the CMD directly 
above 
Region~A, blueward of Region~B.  The apparent magnitudes of these stars 
suggest a distance modulus between 5 and 10
($r\sim0.1-1$~kpc).  In addition, this area of the CMD may also contain
contribution from field blue stragglers and blue horizontal branch stars.

%%%%%%%%%%%%%%%%%%%%%%%%%%%%%%%%%%%%%%%%%%%%%%%%%%%%%%%%%%%%%%%%%%%%%%%%%%%%%%%
%                       REGION___B
%%%%%%%%%%%%%%%%%%%%%%%%%%%%%%%%%%%%%%%%%%%%%%%%%%%%%%%%%%%%%%%%%%%%%%%%%%%%%%%
\subsection{Region B: Galactic Disk F---K Dwarfs, 
  LMC Supergiants} \label{sec:B}

\begin{figure}[p]
\mbox{
\mbox{\epsfysize=8.4cm\epsfbox{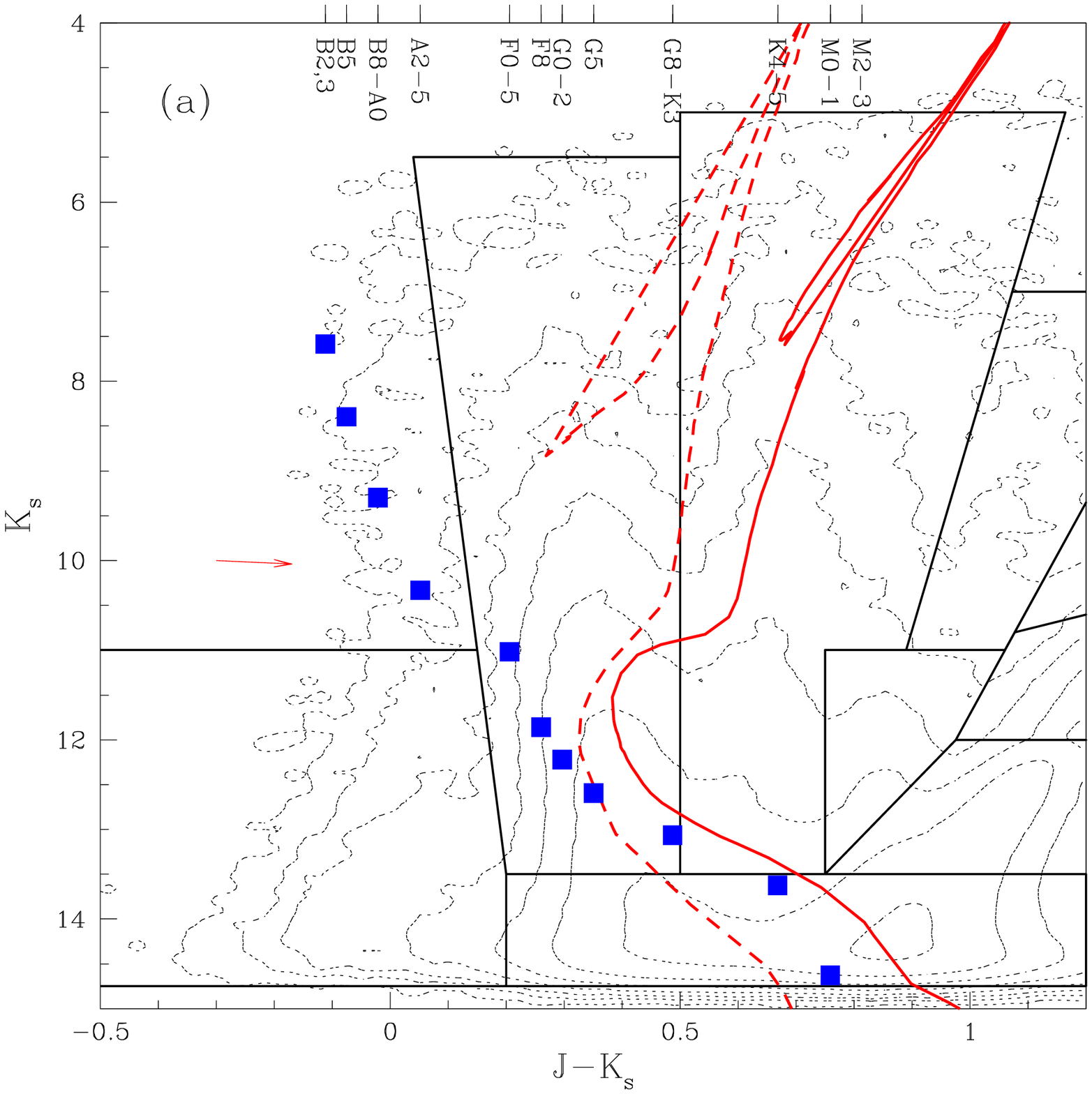}}
\mbox{\epsfysize=8.4cm\epsfbox{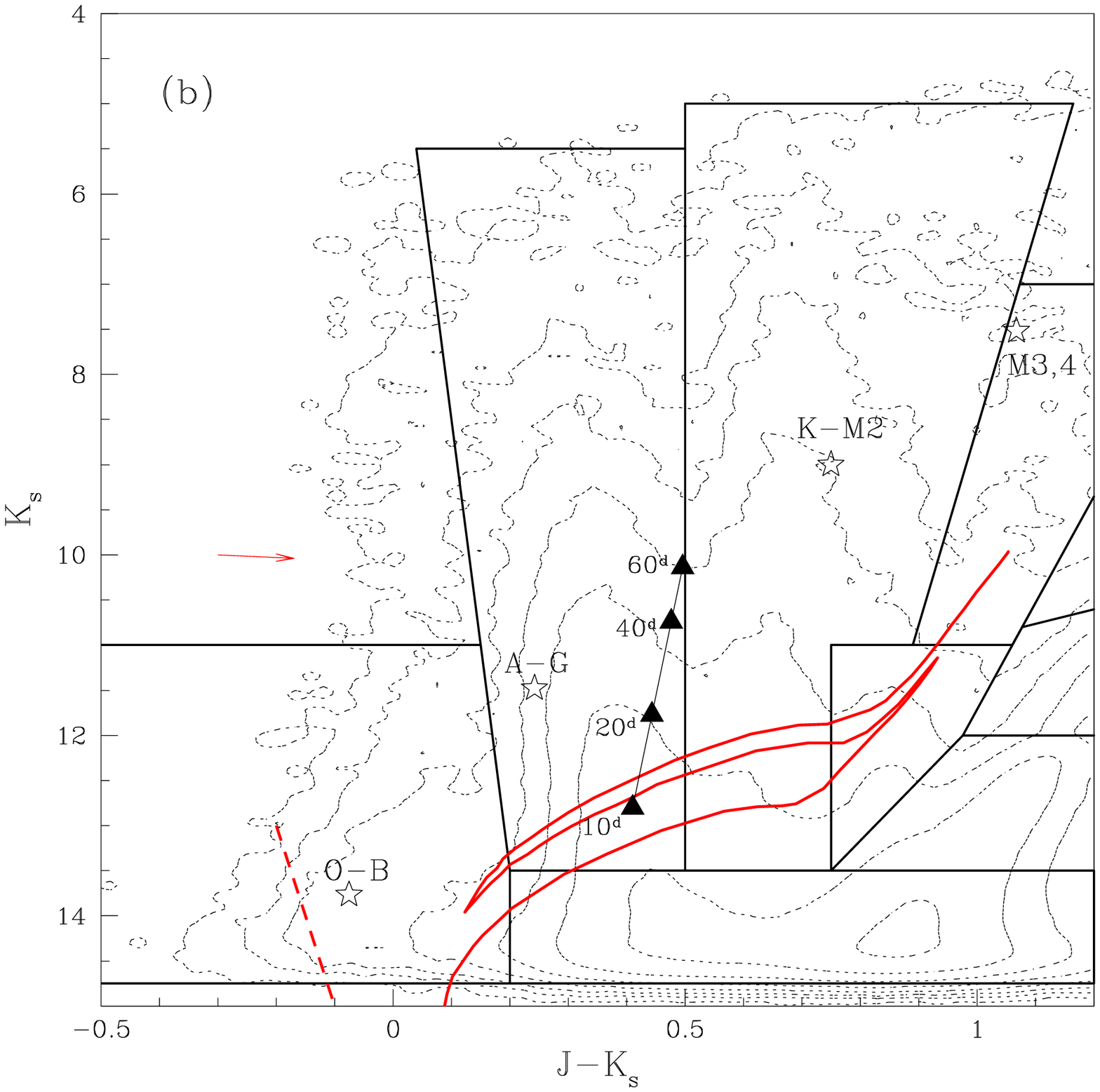}}
}
\caption{Part of the CMD showing Galactic and young LMC populations.
  (a) Galactic dwarf populations in 2MASS CMD.  Solid line: theoretical 
  isochrones for $\tau=7$ Gyr, $Z=0.019$, $\mu=9.0$, $E_{B-V}=0.1$, 
  representing intermediate/old Galactic disk population; dashed line: 
  isochrones for $\tau=14$ Gyr, $Z=0.0004$, $\mu=9.0$, $E_{B-v}=0.1$.  
  Fiducial unreddened dwarf colors from W92 (for $\mu=9.0$) are 
  indicated by filled squares, with the approximate spectral types marked 
  in the top axis.  The reddening vector corresponds to $E_{B-V}=0.2$.
  (b) Young LMC populations.  Solid line shows the theoretical isochrone
  for $\tau=60$ Myr, $Z=0.008$, $\mu=18.5$, $E_{B-V}=0.2$.  Empirical
  colors for Cepheids (Madore \& Freedman 1991) are shown with triangles 
  and mark the location of the instability strip in the diagram.  Stars
  show fiducial color sequence of supergiants (I-II) from W92.  Dashed 
  line shows the tip of the ZAMS at $\mu=18.5$, corresponding to hottest
  O3-O6 dwarfs.  Colors of Cepheids, supergiants and O dwarfs are
  unreddened.  The reddening vector corresponds to $E_{B-V}=0.2$.
\label{fig:match_d}}
\end{figure}

Region~B is a vertically stretched band in the CMD with $J-K_s=0.2-0.5$.  
This color cut isolates the main sequence turnoff of the halo ($J-K_s\approx
0.3$) and the disk ($J-K_s\approx0.4$).  The spatial density distribution 
increases toward NE corner of the field (Fig.~\ref{fig:skyregs}B), i.e. 
toward the Galactic center, and indicates a predominantly Galactic 
population.  The vertical extent in the CMD indicates a wide range of 
distance moduli for these stars.  Based on relative population abundances in 
our synthetic Galactic CMD, we conclude that these sources are disk dwarfs 
of spectral classes in the range from late F to early K.  These stars 
account for $\sim 90\%$ of the foreground source density in the region.  
Their position in the CMD (see Figure~\ref{fig:match_d}a) suggests that the 
dwarfs have distance moduli $\mu=3-10$ ($r\sim 0.04-1.0$ kpc).  Galactic 
giants in the region are of types F---G, but their contribution to 
foreground source density is insignificant, smaller than $5\%$.

\begin{figure}[h]
\epsfysize=9.0cm
\centerline{\epsfbox{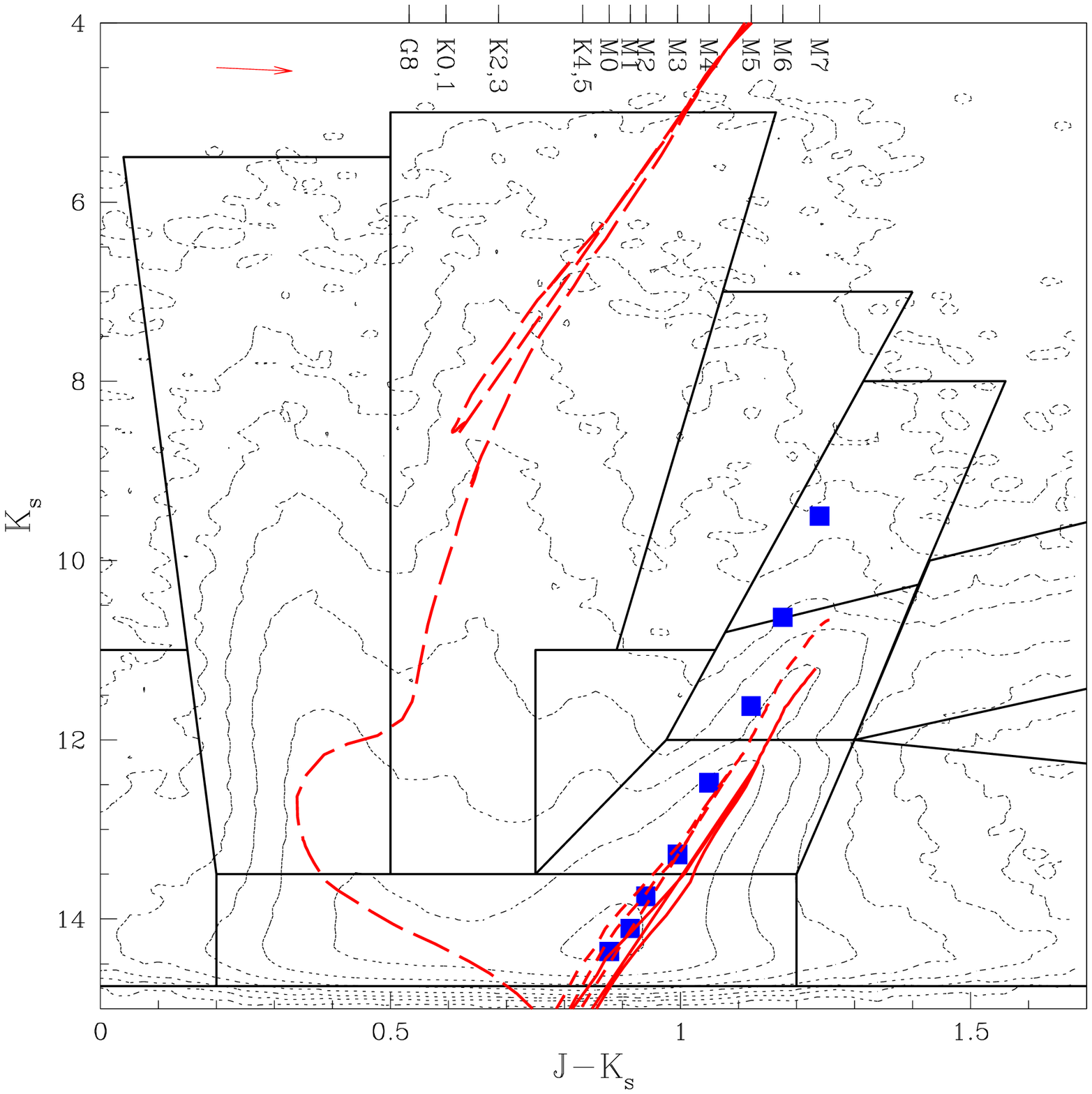}}
\caption{Part of the CMD showing intermediate and old LMC populations.
  Theoretical isochrones represent prototype populations --- solid line: 
  $\tau=11$ Gyr, $Z=0.004$, $\mu=18.5$, $E_{B-V}=0.2$; dashed line: 
  $\tau=4$ Gyr, $Z=0.004$, $\mu=18.5$, $E_{B-V}=0.2$.  The long-dashed
  line shows the isochrone for Galactic RGB stars, $\tau=9$ Gyr, 
  $Z=0.019$, $\mu=10.0$, $E_{B-V}=0.0$.  Fiducial unreddened RGB color 
  sequence (W92) at $\mu=18.5$ is shown with squares, with approximate spectral 
  types along the sequence marked in the top axis.  The reddening vector 
  corresponds to $E_{B-V} = 0.2$.
\label{fig:match_g}}
\end{figure}

The distorted shape of the central isopleth in Figure~\ref{fig:skyregs}B 
suggests presence of the LMC population in this region.  The isopleth 
outlines the structure similar to the one seen in the central regions in 
Figure~\ref{fig:skyregs}A.  Note the overdensity near $\alpha=5^h40^m$,
$\delta=-69^\circ$ and $\alpha=5^h35^m$, $\delta=-67^\circ30'$, marking 
positions of superassociations IV and V, respectively \cite{mar76}.  Based 
on the colors, the LMC component is comprised of young blue and yellow 
supergiants, corresponding to spectral types A---G.  This population 
includes luminous blue variables and short period Cepheids ($P\simlt50^d$).  
Figure~\ref{fig:match_g}b shows the Cepheid sequence based on PL relations 
for LMC Cepheids \cite{mad91}.  Figure~\ref{fig:match_d}b also shows the 
colors of supergiants from W92 (Table 2).  In addition, Region~B contains 
the majority ($\simgt 80\%$) of the known LMC Wolf-Rayet stars. 
Most LMC Wolf-Rayet stars have infrared colors in the range $0<J-K_s<0.5$.  
Their numbers, however, are not significant to produce an observable effect 
in the CMD density.

%\begin{figure}[h]
%\epsfysize=9.0cm
%\centerline{\epsfbox{match_d.ps}}
%\caption{Part of the CMD showing the theoretical color sequences of
%  dwarfs from W92 at $\mu = 0.0$, 2.5, 5.0, 7.5 and 10.  The reddening
%  vector is for $E_{B-V} = 0.1$.  Average Galactic reddening in the
%  direction of the LMC is $E_{B-V} = 0.07$ (Harris \etal{} 1997).
%  \label{fig:match_d}}
%\end{figure}

%%%%%%%%%%%%%%%%%%%%%%%%%%%%%%%%%%%%%%%%%%%%%%%%%%%%%%%%%%%%%%%%%%%%%%%%%%%%%%%
%                       REGION___C
%%%%%%%%%%%%%%%%%%%%%%%%%%%%%%%%%%%%%%%%%%%%%%%%%%%%%%%%%%%%%%%%%%%%%%%%%%%%%%%
\subsection{Region C: Disk K Dwarfs and K Giants, 
  Young Supergiants in the LMC Bar} \label{sec:C}

Similar to B, Region~C is stretched along the magnitude axis, indicating that 
the CMD feature is formed by sources at a range of distances.  The colors of 
this population are in a tight range, $\Delta(J-K_s)\sim0.3$.  Our synthetic 
Galactic CMD suggests that most ($\sim 70\%$) of the observed density in this 
region is produced by disk K dwarfs at $\mu<9$ ($r\simlt 600$~pc).  Disk K 
giants are also present in this region.  Most of them have $\mu\sim 6-13$ 
($r\sim 0.2-4$~kpc).  They contribute $\sim 20\%$ of the foreground density.
The inspection of isochrones in Figure~\ref{fig:match_d}a suggests that Galactic 
giants in this region are in the evolutionary phase of red clump/horizontal
branch stars.  The intrinsic brightness and color of these stars ($M_K=-1.4
\pm0.1$, $J-K_s=0.6\pm0.1$) make them natural candidates in 
this region.  Because of the narrow magnitude range of the clump, the source 
distribution in Region~C along the magnitude axis could help constrain the 
structure of the Galactic disk.

The LMC population in Region~C (seen in Figure~\ref{fig:skyregs}C) is 
slightly older than youngest supergiants in Regions~A, B.  The central isopleths
of the figure outline the bar of the LMC and show no overdensity at the
positions of superassociations, seen in previous two regions.  Most of the 
LMC sources in Region~C have $K_s>10.5$.  The similarity in the shapes of 
central isopleths between Fig.~\ref{fig:skyregs}C and \ref{fig:skyregs}I 
suggests they are lower mass young supergiants with ages $300-500$ Myr, 
evolving into Region~I (Figure~\ref{fig:match_d}a).  These stars trace the 
bar of the Cloud \cite{gre98}.  Some contribution from more massive 
supergiants, including longer-period Cepheids ($P\simlt 100^d$) may 
also be present.

%%%%%%%%%%%%%%%%%%%%%%%%%%%%%%%%%%%%%%%%%%%%%%%%%%%%%%%%%%%%%%%%%%%%%%%%%%%%%%%
%                       REGION___D
%%%%%%%%%%%%%%%%%%%%%%%%%%%%%%%%%%%%%%%%%%%%%%%%%%%%%%%%%%%%%%%%%%%%%%%%%%%%%%%
\subsection{Region D: Disk G---M Dwarfs and LMC RGB and Early AGB Stars} 
  \label{sec:D}

Region~D is the most heavily populated area of the CMD: it includes more than 
a half of all sources in the sample.  Because of its position in the CMD and
the large color range it spans ($0.25<J-K_s<1.2$), this region is also the 
most inhomogeneous.  The spatial distribution, shown in 
Figure~\ref{fig:skyregs}D shows both the foreground and LMC populations (note
the distorting effect of Galactic populations on outer LMC isopleths).  
The observed CMD is distinctly bimodal in this region, with the red half 
populated by RGB and early AGB stars in the LMC, and the blue half populated 
mostly by G---M dwarfs in the Galaxy.  The AGB stars in the red half of the
region ($J-K_s>0.7$) are in their `early-AGB' (E-AGB) phase, during which the energy 
is produced in the thick helium shell and outer hydrogen shell is 
extinguished.  These stars have recently passed the base of the AGB, the 
so-called `AGB-bump' at $K_s\approx16$ that marks the transition from core
to shell helium burning \cite{cas91}.  The AGB-bump was first 
observed by Hardy \etal{} (1984) in their CMD of the LMC bar.  At only 
$\approx 1$ mag brighter than the horizontal branch, this feature is not 
visible in the CMD in Figure~\ref{fig:cmdregs} although it is present in deeper
data (Figure~\ref{fig:deep}).  Empirically, most stars at the E-AGB are 
M type (oxygen-rich).  

Only foreground stars are a minor contributor to the red half of Region~D.
(cf. Figure~\ref{fig:cmds}).  Galactic dwarfs in this region have
$\mu \sim 8-11$ ($r\sim 0.4-1.6$~kpc).  Populations contributing
insignificantly to the source density in this region include young 
supergiants, Cepheids, intermediate mass red stars in the vertically
extended red clump (VRC; see \S\ref{sec:I}).

%%%%%%%%%%%%%%%%%%%%%%%%%%%%%%%%%%%%%%%%%%%%%%%%%%%%%%%%%%%%%%%%%%%%%%%%%%%%%%%
%                       REGION___E
%%%%%%%%%%%%%%%%%%%%%%%%%%%%%%%%%%%%%%%%%%%%%%%%%%%%%%%%%%%%%%%%%%%%%%%%%%%%%%%
\subsection{Region E: Upper RGB and Tip of the RGB} \label{sec:E}

Region~E covers the upper RGB and includes the tip of the RGB (see 
\S\ref{sec:lf}).  Most of these stars are on the first-ascent red giant 
branch; they have degenerate helium cores and hydrogen burning shells.  The 
majority of these stars have ages anywhere between $1$ and $15$ Gyr old.  
The tip of the RGB is defined the helium flash, the ignition of the degenerate 
helium core in old (low-mass) stars (Renzini~\& Fusi~Pecci 1989).  
Stars at the TRGB ignite helium in their cores 
and evolve rapidly to the horizontal branch.  The region also contains 
a significant fraction of AGB stars in transition from E-AGB to
TP-AGB, the stage at which the outer hydrogen shell is re-ignited \cite{ibe83}.  During thermal
pulses, the star begins alternating between hydrogen and helium shell 
burning.  The transition from E-AGB to TP-AGB is theoretically predicted to 
occur near the TRGB.  While on the TP-AGB, these stars may also experience the 
shorter-term atmospheric pulsations that lead to Mira-type variability.  
Analysis of MACHO data \cite{alv98,woo99} suggests that 
essentially all stars brighter and redder than the TRGB are variable.
Most of the E-AGB stars in this region are M stars.  Extrapolated to 
brighter magnitudes, the sequence of oxygen-rich AGB stars extends to 
Regions~F and G (\S\S\ref{sec:F}, \ref{sec:G}).  

Stars in this region of the CMD carry the most weight in our analysis of the 
RGB+AGB luminosity function (see \S\ref{sec:lf}).  Their spatial distribution 
is relatively smooth, showing strong disk and bar components.  Note the 
absence of significant foreground population in Figure~\ref{fig:skyregs}E:
the outer contours are elliptical in shape.  A small fraction of foreground
sources in this region is due to disk M dwarfs.  Their density is steadily 
increasing toward fainter magnitudes (cf. Figure~\ref{fig:cmds}).

%%%%%%%%%%%%%%%%%%%%%%%%%%%%%%%%%%%%%%%%%%%%%%%%%%%%%%%%%%%%%%%%%%%%%%%%%%%%%%%
%                       REGION___F
%%%%%%%%%%%%%%%%%%%%%%%%%%%%%%%%%%%%%%%%%%%%%%%%%%%%%%%%%%%%%%%%%%%%%%%%%%%%%%%
\subsection{Region F: O-Rich AGBs} \label{sec:F}

\begin{figure}[htb!]
\epsfysize=14.0cm
\centerline{\epsfbox{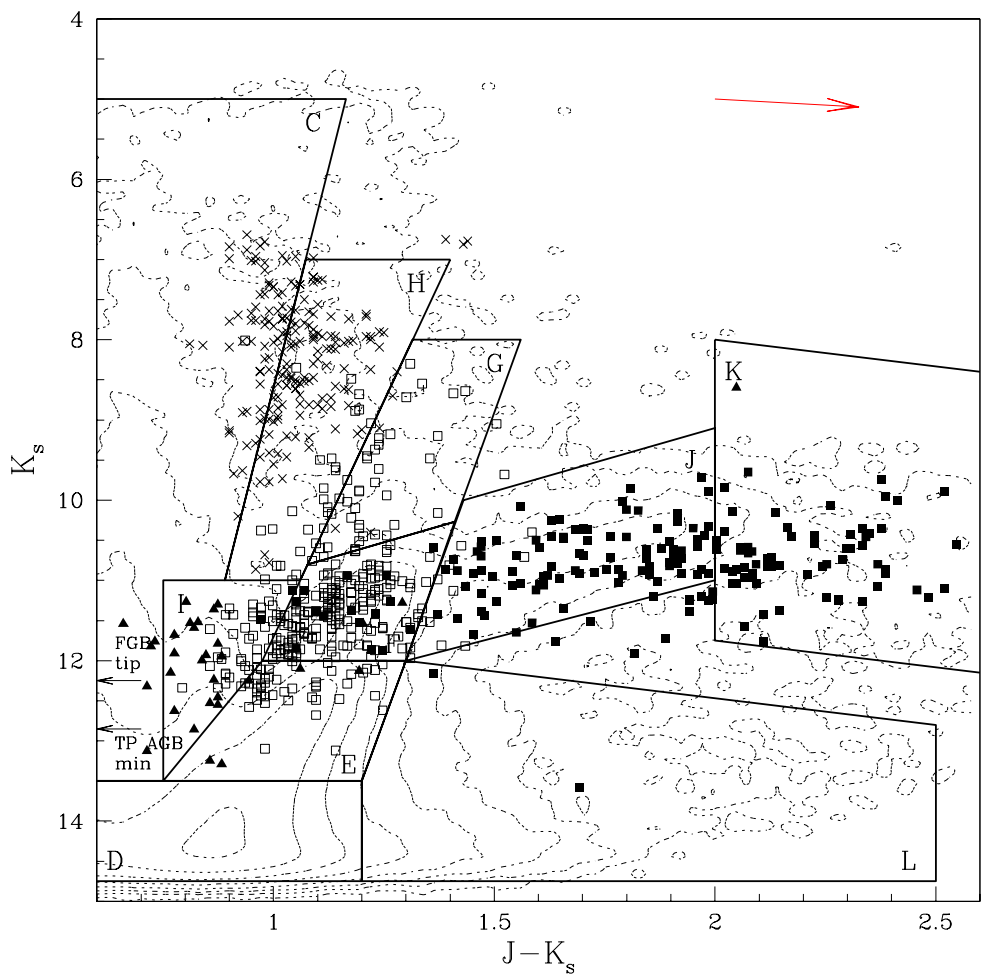}}
\caption{Portion of the CMD showing evolved stars.  Crosses indicate M
  supergiants M1---M4 from Elias \etal{} (1985), solid triangles ---
  K-type LPVs, open squares --- M-type LPVs, and solid squares ---
  C-type LPVs from Hughes \& Wood (1990).  Arrows at left show the
  theoretical tip of the RGB stars and the lower luminosity limit for
  thermally pulsing AGBs.  The reddening vector is for $E_{B-V} =
  0.5$.
\label{fig:miras}}
\end{figure}

Region~F contains primarily oxygen-rich AGB stars of intermediate age
($\simgt 1$ Gyr) that are the descendants of stars in Region~E (note 
the similarity between Fig.~\ref{fig:skyregs}E and \ref{fig:skyregs}F).  
These are E-AGB and TP-AGB stars.  The outer CMD isopleth in this region 
(Figure~\ref{fig:cmdregs}) is distorted and extends into Region~J and
indicates the presence of carbon stars in this region.  During thermal 
pulses, the outer convective envelope may reach into the region where He 
has been transformed into C and bring carbon-enriched material to the 
surface.  This dredge-up process leads to an increase in C/O ratio, and 
M stars in Regions~F and G may become carbon stars. In the CMD, carbon 
stars form a `branch' with redder colors, 
$J-K_s \simgt 1.4$ (see \S\ref{sec:J}). 
Some fraction of Region~F stars are LPVs (see Figure~\ref{fig:miras})
and reddened supergiants.
Figures~\ref{fig:skyregs}F and G do not show isopleths due to a Galactic 
population.  The Galactic component in these regions are negligible because it
is both too bright and too red for disk M dwarfs and too faint for Galactic 
AGB stars.

%%%%%%%%%%%%%%%%%%%%%%%%%%%%%%%%%%%%%%%%%%%%%%%%%%%%%%%%%%%%%%%%%%%%%%%%%%%%%%%
%                       REGION___G
%%%%%%%%%%%%%%%%%%%%%%%%%%%%%%%%%%%%%%%%%%%%%%%%%%%%%%%%%%%%%%%%%%%%%%%%%%%%%%%
\subsection{Region G: AGB Stars} \label{sec:G}

Region~G contains the most massive stars with degenerate C/O cores.
This is a population of young AGB, post core-helium burning stars 
with initial masses between about 5-8 solar masses.  These are too short-lived to 
become carbon stars ($0.1-1$ Gyr old), but not massive enough to become red supergiants.  
Similar to Region~F,
this region also includes LPVs (Figure~\ref{fig:miras}).
The period-color relation for oxygen-rich Miras derived from Feast 
\etal{} (1989) rather closely traces the young AGB branch of the CMD
outlined by Regions~F and G \cite{wei00}.  Region~G 
lies at bright 
enough apparent magnitudes that the foreground density of M dwarfs is 
low, even in comparison to the relatively small number of O-rich 
luminous AGB stars in the LMC.  

%%%%%%%%%%%%%%%%%%%%%%%%%%%%%%%%%%%%%%%%%%%%%%%%%%%%%%%%%%%%%%%%%%%%%%%%%%%%%%%
%                       REGION___H
%%%%%%%%%%%%%%%%%%%%%%%%%%%%%%%%%%%%%%%%%%%%%%%%%%%%%%%%%%%%%%%%%%%%%%%%%%%%%%%
\subsection{Region H: LMC K---M Supergiants, Galactic M Dwarfs, 
  K---M Giants} \label{sec:H}

Panel H of Figure~\ref{fig:skyregs} reveals an LMC population.  The
spatial distribution is similar to the distribution of young OB stars
(Region~A), suggesting that these objects are also relatively young.
At $\mu = 18.5$, these stars are too bright to be normal M giants.  Based on
their near infrared colors, we identify the Region~H sources as supergiants of 
M type.  They trace the spiral structure of the LMC and do not show 
significant overdensity in the bar of the Cloud, consistent with a young 
population.  The masses of these stars are believed to be
$\sim 2 - 9$ solar masses \cite{ber85}.  In the evolutionary 
sequence, these stars are descendants of stars in Region~A (note the 
similarity between corresponding panels in Figure~\ref{fig:skyregs}).  
These stars are also the high mass extension of the VRC.

In Figure~\ref{fig:miras}, we plot the observed colors of M1---M4 
supergiants from the sample of Elias \etal{} (1985).  The
colors of their sample occupy a portion of Region~H, 
supporting our identification.  The Galactic foreground consists of 
roughly equal contributions from disk K---M giants and M dwarfs, but 
their overall contribution to the source density in the region is only 
a few percent.  This is confirmed by the absence of the Galactic 
isopleths in Figure~\ref{fig:skyregs}H.

%%%%%%%%%%%%%%%%%%%%%%%%%%%%%%%%%%%%%%%%%%%%%%%%%%%%%%%%%%%%%%%%%%%%%%%%%%%%%%%
%                       REGION___I
%%%%%%%%%%%%%%%%%%%%%%%%%%%%%%%%%%%%%%%%%%%%%%%%%%%%%%%%%%%%%%%%%%%%%%%%%%%%%%%
\subsection{Region I: LMC Intermediate-Mass Red Supergiants, 
  Galactic K---M Dwarfs} \label{sec:I}

Region~I is located at the center of the CMD, at $0.7\le J-K_s\le 1.0$.  
The observed overdensity in the CMD is associated with the vertically 
extended red clump (Zaritsky \& Lin 1997).  This feature consists of
intermediate mass stars and is the low-mass extension 
of the red supergiants (Region~H).  The VRC extends upward from the red 
clump at $K_s\approx17$ and becomes visible in the CMD
near $K_s=13.5$.  At the this point, the redward slope of the RGB is
sufficient to distinguish the VRC.

The spatial distribution is dominated by the bar and shows traces of the 
spiral structure.  We conclude that this LMC population is young, with the 
age $\simlt 500$ Myr.  The major LMC contributors to the source density
in this region are K and M supergiants.  This is supported by the 
overall similarity of LMC isopleths in Figures~\ref{fig:skyregs}I and
\ref{fig:skyregs}H, and also the fact that Region~I is at the extension 
of Region~H to fainter magnitudes  and lower masses.  
Figure~\ref{fig:miras} shows K and 
M type Miras and SR variables in the sample of Hughes \& Wood (1990).
A significant fraction of their variables falls in this CMD region
suggesting that some of these 2MASS stars 
also are variables.

The distribution in Figure~\ref{fig:skyregs}I also reveals foreground
populations.  The Galactic foreground consists of M and late K dwarfs.
Galactic giants contribute less than 5\% of the foreground density.  
The dwarfs are located in the disk, with distance moduli $\mu=5-8$ ($r\sim 
0.1-0.4$~kpc; Figure~\ref{fig:match_d}).  The contribution from the 
Milky Way halo is smaller than 0.0005\% by number for this and for all other 
regions of the CMD.

%%%%%%%%%%%%%%%%%%%%%%%%%%%%%%%%%%%%%%%%%%%%%%%%%%%%%%%%%%%%%%%%%%%%%%%%%%%%%%%
%                       REGION___J
%%%%%%%%%%%%%%%%%%%%%%%%%%%%%%%%%%%%%%%%%%%%%%%%%%%%%%%%%%%%%%%%%%%%%%%%%%%%%%%
\subsection{Region J: Carbon Stars in the LMC} \label{sec:J}

At $J-K_s \simgt 1.4$, Region~J sources are primarily carbon-rich
TP-AGB stars.  These stars are descendants of oxygen-rich TP-AGBs in
Regions~F and G.  Their outer layers are enriched in C through
convection from stellar interior.  As mentioned in \S\ref{sec:E}, most
of these stars are long-period variables.  The variability cannot be
determined based on single epoch 2MASS data, but the well-defined
sequence motivates a follow-up campaign.  Figure~\ref{fig:miras} shows
the sample of C-rich LPVs from Hughes \& Wood (1990) overplotted on
the 2MASS CMD.  The contamination by M-type LPVs is small.  The
spatial distribution of C stars in the field is similar to the
distribution of their precursors (Figs.~\ref{fig:skyregs}F).  The
distribution is rather smooth and shows a loop of stellar
material\footnote{The feature to the SE of the bar in
  Figure~\ref{fig:skyregs}J, near $\alpha\approx5^h50^m$,
  $\delta\approx-73^\circ$ represents a hole in the disk.}, which has
been described by Westerlund (1964).  The loop is the extension of the
main northern spiral arm circling the main body of the system and
returning toward the bar after a nearly complete turn.

Sources in this region of the CMD offer the best opportunity to study the
three-dimensional structure of the LMC for two reasons.  First, the spatial 
coverage of the Cloud achieved by 2MASS is total and allows to probe the 
entire LMC.  Second, as long-period variables, Region~J stars are 
potentially good standard candles, since their intrinsic luminosity can be 
characterized based on their period or color.  Given the selection 
efficiency\footnote{owing to their extremely red colors, these stars are 
uncontaminated by other populations} and easily quantifiable intrinsic 
brightness through the period-luminosity-color relation (e.g., Feast \etal{} 
1989), these stars are excellent probes of the LMC structure along the line 
of sight.  Preliminary results (WN) indicate 
that the width of the 
intrinsic brightness distribution is smaller than $\sigma_M = 0.2$
magnitudes in a narrow color range, $\Delta (J-K_s) \sim 0.1$.
At this accuracy, these standard candles can resolve features in the LMC at
$\Delta r \sim 4.5$ kpc.  2MASS detected approximately 
$10^4$ potential carbon LPVs and these are sufficient to attain a reasonable confidence 
level in the inferred spatial structure.  In Weinberg \& Nikolaev (2000), 
we present our study of the three-dimensional structure of the LMC.

%%%%%%%%%%%%%%%%%%%%%%%%%%%%%%%%%%%%%%%%%%%%%%%%%%%%%%%%%%%%%%%%%%%%%%%%%%%%%%%
%                       REGION___K
%%%%%%%%%%%%%%%%%%%%%%%%%%%%%%%%%%%%%%%%%%%%%%%%%%%%%%%%%%%%%%%%%%%%%%%%%%%%%%%
\subsection{Region K: Dusty AGBs} \label{sec:K}

An extension of Region~J, Region~K contains extremely red objects.  We 
identify them with obscured AGB carbon-rich stars.  Their large $J-K_s$ 
colors are due to dusty circumstellar envelopes ($E_{B-V}\simgt 1$).
The latter is confirmed by the appearance of their spatial 
and CMD distributions: (1)~Figure~\ref{fig:skyregs}K shows traces of the 
spiral structure outlined by these sources; and (2)~the 
distribution in the CMD spreads from the end of Region~J in the direction 
of reddening vector.  Matching with existing near infrared photometry of obscured 
AGB stars in the LMC \cite{zij96,vln98} shows that most of these 
sources are indeed in this region of the CMD.  Other extremely red 
populations could also be found here, e.g. `cocoon' stars \cite{rei91}, 
or OH/IR stars \cite{woo92,vln98}.  In addition, two of the known LMC 
protostars, N159-P1 and N159-P2 \cite{jon86} also fall in this region.

%%%%%%%%%%%%%%%%%%%%%%%%%%%%%%%%%%%%%%%%%%%%%%%%%%%%%%%%%%%%%%%%%%%%%%%%%%%%%%%
%                       REGION___L
%%%%%%%%%%%%%%%%%%%%%%%%%%%%%%%%%%%%%%%%%%%%%%%%%%%%%%%%%%%%%%%%%%%%%%%%%%%%%%%
\subsection{Region L: Reddened LMC M Giants, Galactic M Dwarfs and
  2MASS Galaxies} \label{sec:L}
 
Stellar sources in Region~L are reddened M giants in the LMC and a
small number of reddened Galactic M dwarfs.  However, a significant
number of sources in this region are background galaxies.  The
predicted CMD density in Region~L due to Galactic stars is too low
even after the decrease in the photometric quality near the flux limit
has been taken into account; the decreasing signal-to-noise ratio
causes the apparent widening of the contour levels (see
Figure~\ref{fig:cmds}).  According to Jarrett (1998), more than $90\%$
of 2MASS galaxies have colors redder than $J-K_s=1$.

The spatial distribution of sources (Figure~\ref{fig:skyregs}L) shows
some overdensity toward the center of the LMC.  The densest part of
the diagram corresponds to the position of 30 Doradus complex.  Traces
of spiral structure of the LMC are also visible.  Based on their
colors, these sources are heavily obscured RGB stars in the LMC: they
lie in the direction of the reddening vector from the RGB.  The
inferred reddening for these sources, $E_{B-V} \sim 0.5$, is
consistent with the extended tail of the LMC reddening distribution
\cite{har97}.  Region~L also includes contribution from massive ($>10
M_\odot$) protostars and ultra-compact H II regions, see
Figure~\ref{fig:miras}.

A population of dwarfs is implied by the outer isopleths of 
Figure~\ref{fig:skyregs}L that show the increase in the direction of 
Galactic center.  These are local M dwarfs in the disk of the Milky Way, 
with $\mu \sim 5-8$ ($r \sim 100-400$ pc).  

%%%%%%%%%%%%%%%%%%%%%%%%%%%%%%%%%%%%%%%%%%%%%%%%%%%%%%%%%%%%%%%%%%%%%%%%%%%%%%%
%                       LUMINOSITY   FUNCTION
%%%%%%%%%%%%%%%%%%%%%%%%%%%%%%%%%%%%%%%%%%%%%%%%%%%%%%%%%%%%%%%%%%%%%%%%%%%%%%%
\section{Luminosity Function of LMC RGB and AGB Populations} \label{sec:lf}

\begin{figure*}[p]
\centerline{
\mbox{\epsfysize=8.3cm\epsfbox{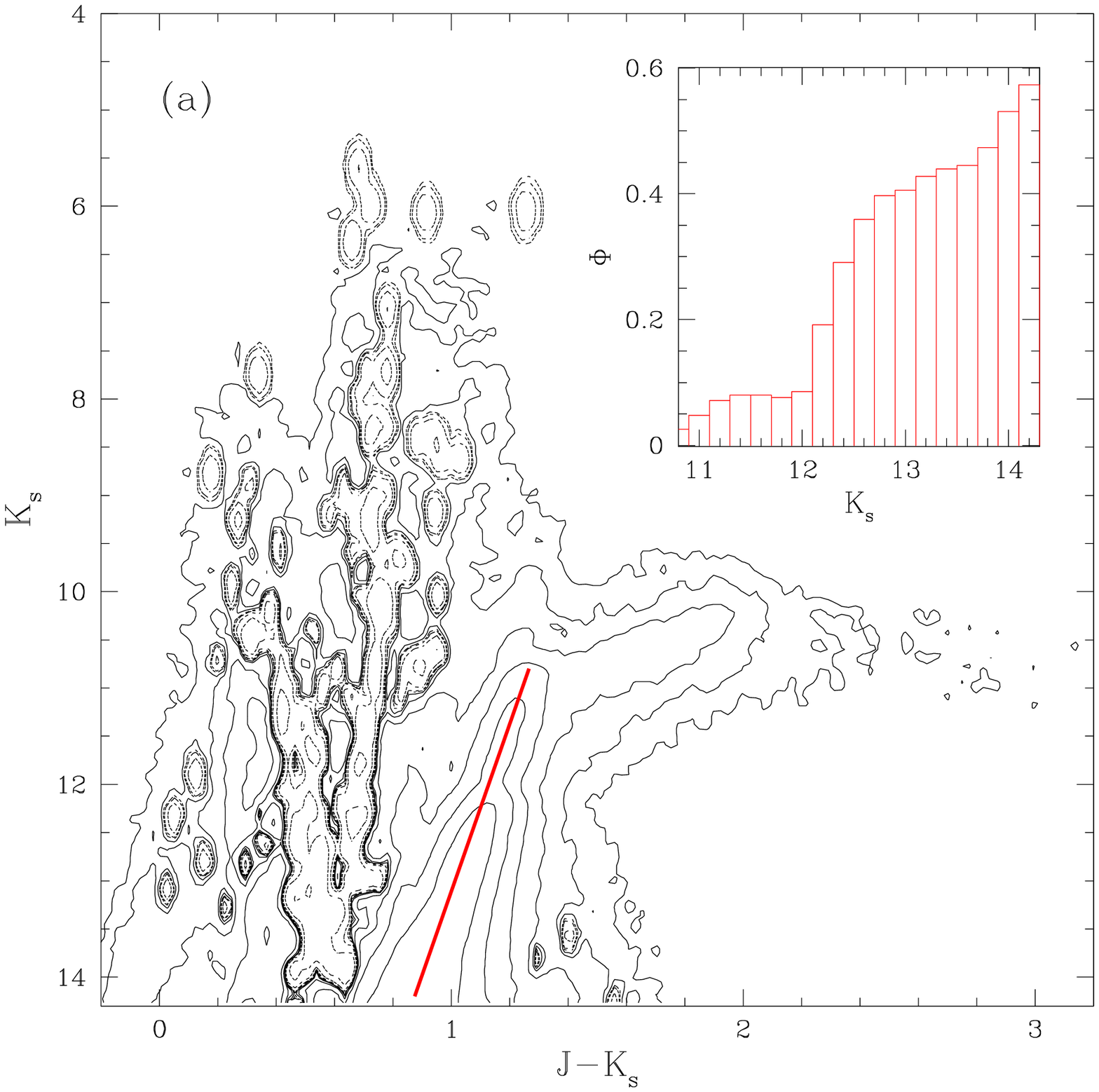}}
}
\mbox{
\mbox{\epsfysize=8.3cm\epsfbox{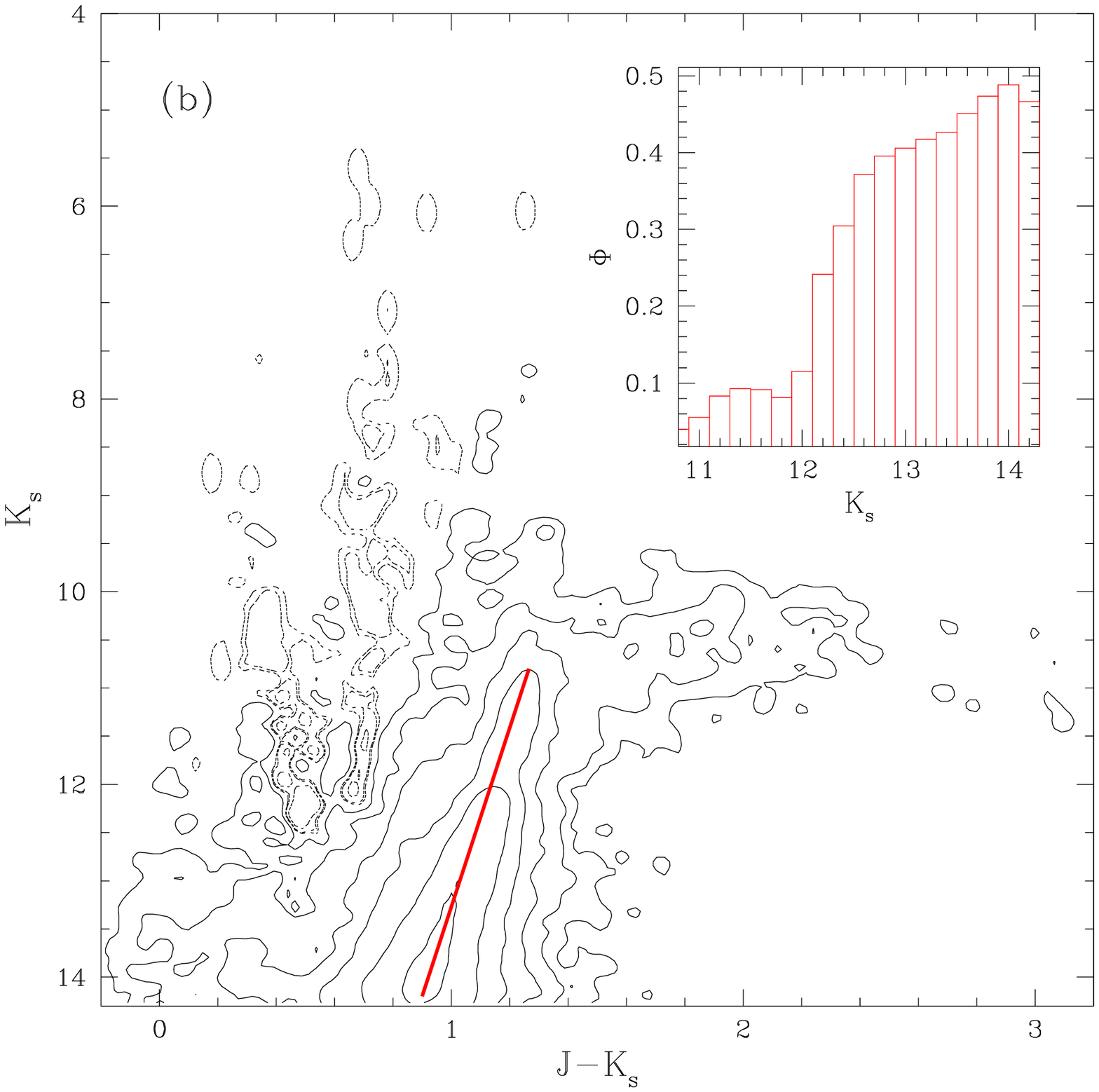}}
\mbox{\epsfysize=8.3cm\epsfbox{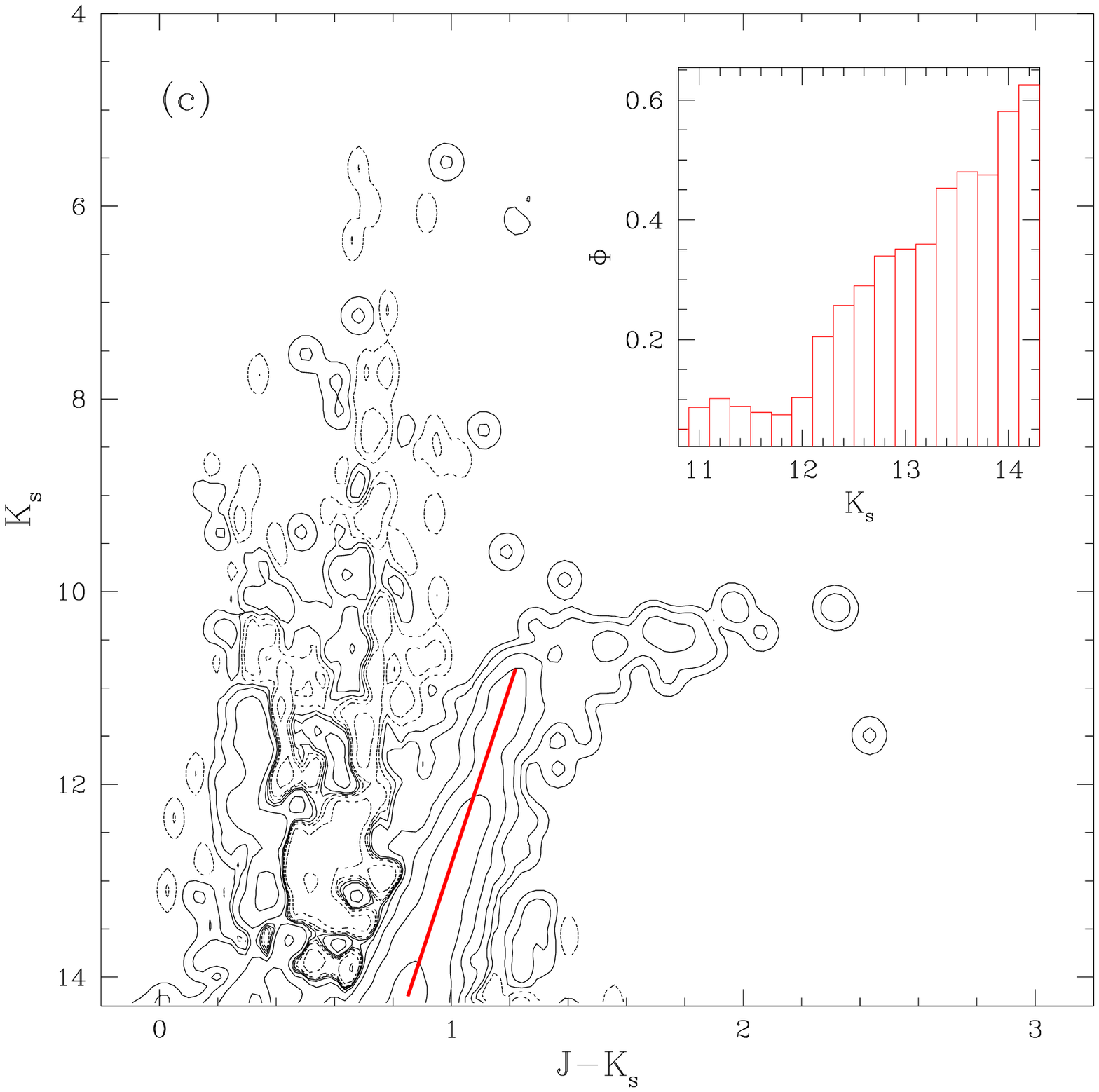}}
}
\caption{Color-magnitude diagrams of the LMC populations and the apparent
  luminosity functions of the LMC giants.  Panels show CMDs of (a)
  entire field, (b) bar field, and (c) outer loop field (see text).
  The CMDs are constructed by subtracting Galactic foreground
  contribution from CMD in Figure~\ref{fig:cmdregs}, normalized to the
  same sky area.  Contour levels in all panels are logarithmic, spaced by
  0.5, from 2.5 to 5.5 (a), 2.0 to 4.0 (b), and from 1.0 to 3.5 (c).
  Dotted lines indicate negative density regions.  Luminosity functions,
  normalized to unity, are shown in the insets of each panel.
\label{fig:lfs}
}
\end{figure*}

We derive the LMC giant branch luminosity function (LF) from the color-magnitude 
diagram after subtracting Galactic foreground.  Since we did not have the 
access to Galactic data at the time, the foreground contribution was 
estimated from three small fields located at the edges of our LMC field.
We then scaled the resulting foreground CMD to the 
entire LMC field by using the estimate for the number of Galactic sources
from our synthetic model.  Figure~\ref{fig:lfs}a shows the field CMD 
for LMC populations only, after subtracting Galactic foreground.  The 
expected Galactic source counts is $\sim 4\times10^5$ 
or about $50\%$.  The uncertainties in the observed CMD
and in the Galactic model result in negative density regions (dotted
contours) in Figure~\ref{fig:lfs}.  The average negative density
is $-10^{3.2}$ mag$^{-2}$.

The luminosity function of the LMC giants is obtained by projecting
the color-magnitude diagram perpendicular to the giant branch ridge
line.  The function is normalized to unity.  Numerical values
for the RGB luminosity function are given in Table~\ref{table:lf} and
shown in the inset to Figure~\ref{fig:lfs}.  A strong feature of the
LF is a significant excess at $K_s \approx 12.5$, due to the TRGB.
From the analysis of the derivative of the apparent luminosity
function, we derive the position of the TRGB at $K_s=12.3\pm0.1$.
Brightward of the TRGB, the number of RGB stars drops off.  The
increase in the number density at the faint end, $K_s\simgt14$, is due
to the increased contribution from Galactic M dwarfs (cf.
Fig.\ref{fig:cmds}).  At the bright end of the magnitude range,
$11<K_s<12$, the luminosity function is nearly constant.  It is also
well above the expected number from extrapolated RGB counts.  The
fraction of RGB stars is small in this magnitude range and most of the
stars contributing to the LF are on the AGB.  As discussed above
(\S\ref{sec:F}), these stars tend to be oxygen-rich, but carbon-rich
AGBs (and LPVs) are also present.

We select two $2^\circ \times 1^\circ$ fields, one near the optical 
center of the bar, and the other one at $\alpha = 93^\circ$, 
$\delta=-67.5^\circ$ (J2000.0), near the outer loop, to compare the 
observed M giants luminosity functions in distinct LMC environments.
The outer field is probing the LMC's outer loop delineated by
the evolved stars (see \S\ref{sec:J}).  For each of the fields, we
subtract the estimated Galactic foreground density scaled to the surface 
area of the fields.  Similar to Fig.~\ref{fig:lfs}a, inaccuracies in both
observations and models produce negative density regions.  However,
the average negative density in these regions is small compared
to the giant branch, only $-10^{0.4}$ mag$^{-2}$ for bar region and 
$-10^{0.5}$ mag$^{-2}$ for loop field.  
Comparing the two CMDs qualitatively, we note that contribution of 
young OB stars and supergiants, at $J-K_s\simlt0.2$, appears stronger in 
the central regions of the LMC.  Even though the luminosity functions in 
the respective fields appear different, careful analysis shows that
the difference is superficial: 
$\chi^2/$d.o.f.$\approx 0.2$, which gives no motivation 
to entertain any difference in the parent populations.  The bar 
LF is similar to the luminosity function of the entire field 
(Figure~\ref{fig:lfs}a), which is not surprising because the source 
density in the LMC field is dominated by the LMC bar.  The bar luminosity 
function also has a pronounced TRGB at $K_s\approx12.3$ and excess
number density due to AGBs at $11<K_s<12$.  The sharp increase due 
to Galactic M stars, seen in Figure~\ref{fig:lfs}a at $K_s\simgt14$, has
disappeared in the bar field, because we have boosted the
ratio of LMC/Galactic counts by narrowing 
down the field to the area of greatest LMC density.  The off-bar LF shows
only a mild increase in the source counts at the location of TRGB, but has
the same, roughly constant profile at $K_s<12$, due to the AGB population,
visible in the other two luminosity functions.  
To quantify both LFs, we present their numerical
values in Table~\ref{table:lf}.  The luminosity functions are given in
relative units, normalized to unity.  The table also gives the source
counts for the LMC giant branch.  For the entire LMC field we present
the total counts per magnitude bin, and for the two smaller fields we
give stellar density (counts~mag$^{-1}$~deg$^{-2}$).

\begin{table}[t!]
\centering
\caption{Apparent luminosity function and source number density for the
  LMC giants}
\label{table:lf}
\bigskip
\begin{tabular} {c|ccc|ccc}
  \hline
  $K_s$ & \multicolumn{3}{c|}{$\Phi$ (mag$^{-1}$)$^{\rm a}$} &
  \multicolumn{3}{c}{\parbox[t]{3.5cm}{log number density (mag$^{-1}$ deg$^{-2}$)}} \\
  & LMC & Bar & Loop & LMC$^{\rm b}$ & Bar & Loop \\
  \hline
 10.8 &   0.03 &   0.04 &   0.05 & 3.16 & 2.17 & 1.16 \\
 11.0 &   0.05 &   0.06 &   0.09 & 3.44 & 2.32 & 1.39 \\
 11.2 &   0.07 &   0.08 &   0.10 & 3.61 & 2.50 & 1.46 \\
 11.4 &   0.08 &   0.09 &   0.09 & 3.65 & 2.54 & 1.40 \\
 11.6 &   0.08 &   0.09 &   0.08 & 3.66 & 2.54 & 1.35 \\
 11.8 &   0.08 &   0.08 &   0.08 & 3.64 & 2.48 & 1.33 \\
 12.0 &   0.09 &   0.12 &   0.11 & 3.68 & 2.63 & 1.47 \\
 12.2 &   0.19 &   0.24 &   0.21 & 4.04 & 2.96 & 1.76 \\
 12.4 &   0.29 &   0.31 &   0.26 & 4.22 & 3.06 & 1.87 \\
 12.6 &   0.36 &   0.37 &   0.30 & 4.31 & 3.14 & 1.92 \\
 12.8 &   0.40 &   0.40 &   0.35 & 4.35 & 3.17 & 1.99 \\
 13.0 &   0.40 &   0.41 &   0.36 & 4.36 & 3.18 & 2.00 \\
 13.2 &   0.43 &   0.42 &   0.36 & 4.39 & 3.20 & 2.01 \\
 13.4 &   0.44 &   0.43 &   0.45 & 4.39 & 3.20 & 2.11 \\
 13.6 &   0.44 &   0.45 &   0.47 & 4.40 & 3.23 & 2.14 \\
 13.8 &   0.47 &   0.47 &   0.46 & 4.43 & 3.25 & 2.13 \\
 14.0 &   0.54 &   0.49 &   0.57 & 4.48 & 3.26 & 2.22 \\
 14.2 &   0.58 &   0.47 &   0.61 & 4.51 & 3.24 & 2.25 \\
\hline
\end{tabular}
\par\parbox[t]{12cm}{%
\vskip 0.1cm
$^{\rm a}$The luminosity functions are given in relative counts per magnitude.  

$^{\rm b}$Number density for the entire field is given as source counts per 
magnitude bin.}
\end{table}

We fit theoretical isochrones \cite{gir00} to each giant branch in
Figure~\ref{fig:lfs} to test for differences in metallicity between central 
and outer parts of the LMC.  We chose $20$ equally-spaced grid points in
$K_s$ magnitude between $14.3$ and $12.3$ and compute the peak in the
distribution in $J-K_s$ at these fixed $K_s$ points.
The difference between an isochrone (model) and the RGB (data) 
is characterized by the cost function (mean integrated square error):
\begin{equation}
f = \sum _j [(J-K_s)_{j,\,RGB}-(J-K_s)_{j,\,iso}]^2 + w\, [K_s^{TRGB} - 12.3]^2,
\label{eq:cost}
\end{equation}
where the second term is weighted measure of the match between theoretical
magnitude at helium flash and observed TRGB.  The weight $w$ is an adjustable
parameter on the order of unity.  The cost function (\ref{eq:cost})
is minimized on a grid of parameter values, where the free parameters are 
the log-age $\tau$, metallicity $Z$, distance modulus $\mu$ and average 
reddening $E_{B-V}$.  The best fit isochrones are as follows:
$(\tau, Z, \mu, E_{B-V})=(9.8\pm0.3, 0.004^{+0.002}_{-0.001}, 18.45\pm0.11, 
0.21\pm0.07)$ for the central field and $(\tau, Z, \mu, E_{B-V})=(9.8\pm0.4, 
0.004^{+0.002}_{-0.001}, 18.50\pm0.13, 0.13\pm0.09)$ for the outer field 
(all errors statistical).  These results imply an age range for RGB 
populations from $3$ to $13$ Gyr with an average of $6$ Gyr.  The
slope
degeneracy of the isochrones in the RGB makes specific tests
of star formation history difficult.
In particular, our preliminary RGB isochrone analysis cannot
distinguish between a continuous and single/multiple burst star formation
history of the Cloud prior to approximately $4$ Gyr ago.  
Overall, our results do not indicate a radial metallicity gradient
and provide only
marginal evidence for larger reddening in central fields.  The absence of
strong metallicity gradients in the LMC is in agreement with results of 
Olszewski \etal{} (1991), who found no evidence for abundance gradient for 
cluster system.  Constant C/M star ratio across the face of the LMC 
\cite{wes97} and Cepheid abundances \cite{har83} also support this
result.  Our results imply the range of abundances for field populations
$-0.8<$[Fe/H]$<-0.5$, which is in good agreement with the mean abundance
[Fe/H]$=-0.58\pm0.05$ (systematic) $\pm0.30$ (statistical) for the inner
LMC disk \cite{col99}.  Our results agree with the disk abundance
[Fe/H]$=-0.7$ \cite{cow82}, and with results of Bica \etal{} (1998), who 
derived the range 
$-1.1<$[Fe/H]$<-0.4$ with the average $<[$Fe/H$]>= -0.61\pm0.11$ from fields 
in the outer disk of the LMC.

\begin{figure}[h]
\epsfysize=9.0cm
\centerline{\epsfbox{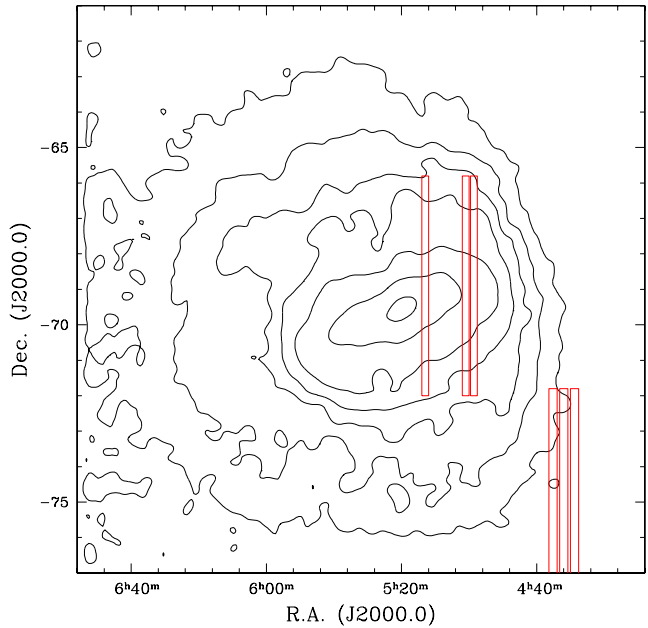}}
\caption{Positions of six `deep' scans in the sky.  Each scan has six times
  the standard integration time.
\label{fig:deep}}
\end{figure}

\begin{figure*}[p]
\epsfysize=18.0cm
\centerline{\epsfbox{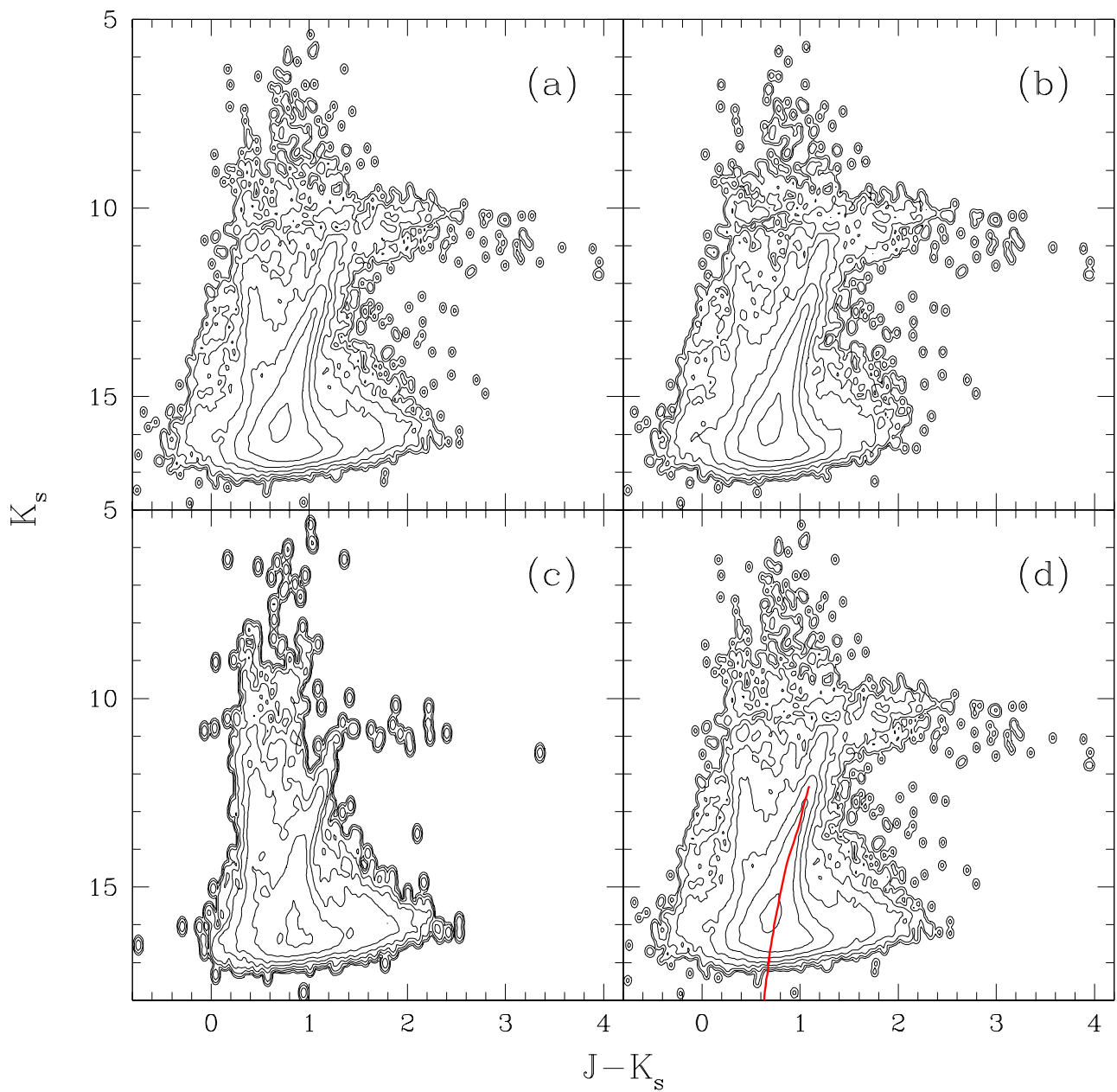}}
\caption{`Deep' color-magnitude diagrams of the LMC.  Panels show CMDs
  of (a) all six deep scans combined, (b) three bar scans, and (c) three
  outer field scans.  Panel (d) is the same as panel (a), except also   
  shows the best-fit isochrone for the RGB.  Galactic foreground is not
  subtracted.  The contour levels are spaced logarithmically by 0.5, 
  from $-3.6$ to $-0.1$ (a,b,d), and from $-3.7$ to $-0.2$ (c).  
  The lower RGB is enhanced as compared to Figure~9.  Weakness of RGB 
  in panel (c) indicates strong relative contribution from Galactic 
  foreground.  The red giant clump is just below the completeness limit 
  in this diagram, at $K_s\approx17$, $J-K_s\approx0.65$.
\label{fig:deepcmd}}
\end{figure*}

The adopted detection threshold (SNR $= 10$; \S\ref{sec:data}) leads 
to the effective completeness limit of $K_s\approx14.3$ in our data (cf.
Figure~\ref{fig:cmdregs}).  With this flux level,
only the upper RGB is visible, leaving out both the AGB-bump at $K_s\approx16$ 
and the red clump at $K_s\approx17$.  To resolve the giant branch down
to $K_s\approx16-17$, we use 2MASS engineering data, which includes six LMC
scans, positioned as shown in Figure~\ref{fig:deep}.  Each of the `deep' scans 
has six times the standard exposure.  
The color-magnitude diagram of the deep data is shown in Figure~\ref{fig:deepcmd}.
Total number of sources in the diagram is $87,093$, of which $69,878$ ($\sim80\%$)
are in the three bar scans.  Because of the bar dominance in deep data, the CMDs 
of the entire deep sample and the bar scans (Figs.~\ref{fig:deepcmd}a and 
\ref{fig:deepcmd}b, respectively) are similar.  
% The RGB is rather 
% weak in the outer field, Figure~\ref{fig:deepcmd}c, which suggests significant
% relative contribution from Galactic foreground.  
The increased sensitivity reaches
the AGB-bump at $J-K_s=0.7$, $K_s=15.8$, but still shy of the
red giant clump.  As with the main dataset,
we quantify the deep RGB population by fitting isochrones.  The resulting best 
fit parameters are $Z=0.004^{+0.002}_{-0.001}$, $\tau=9.7\pm0.3$,
$\mu=18.50\pm0.12$, $E_{B-V}=0.19\pm0.08$.  The uncertainties here are
statistical errors, derived from the shape of the $\chi^2$ surface near
minimum.  The range of ages for RGB populations inferred from the deep data is similar
to that derived for the main data set: from $3$ to $10$ Gyr, with the average 
of $5$ Gyr.  The other parameters are also consistent with the values derived 
from the regular 2MASS data.  These estimates are in good agreement with recent
results in the literature, e.g., the average reddening $E_{B-V}=0.20$ \cite{har97}; 
the LMC distance from Key Project, $\mu = 18.5$ \cite{mou99}; the LMC distance 
from TRGB, $\mu=18.59\pm0.09$ \cite{sak99}.

\section{Summary} \label{sec:summary}

In this paper, we analyzed the near-infrared CMD of the Large
Magellanic Cloud and identified the major stellar populations.
The populations are identified based on isochrone fitting and
matching the theoretical CMD colors of known populations to the observed 
CMD source density.  Tables~\ref{table:regions} and \ref{table:populations} 
summarize the contents of the CMD regions.
\begin{table*}[t]
\caption{Regions of interest in the 2MASS CMD of the Large Magellanic Cloud.
\label{table:regions}}
\begin{tabular} {ccccr}
\hline
Region & $N_{src}$ & $f_{Gal}$$^{\rm a}$ & Region boundaries &
Dominant Spectral Types$^{\rm b}$ \\
\hline
A &   6,659 & $0.15$   & $11<K_s<14.75$, $J-K_s<0.2$ & {\bf B-A I-II, O3-O6 V}\\
B &  77,204 & $0.80$ & $5.5<K_s<13.5$, $0.2<J-K_s<0.5$ & F-K V \\
C &  62,713 & $0.80$ & $5<K_s<13.5$, $0.5<J-K_s\simlt0.8$ & K V, K III \\
D & 440,472 & $0.45$ & $13.5<K_s<14.75$, $0.2<J-K_s<1.2$ & {\bf K-M III}, F-M V \\
E & 166,263 & $0.05$ & $12<K_s<13.5$, $0.9\simlt J-K_s\simlt1.2$ & {\bf M III}, M V \\
F &  22,134 & $0$            & $10.5\simlt K_s<12$, $1\simlt J-K_s\simlt1.3$ & {\bf M, MS} \\
G &   1,438 & $0$            & $8<K_s\simlt10.5$, $1.2\simlt J-K_s\simlt1.5$ & {\bf M, MS} \\
H &   2,450 & $0.05$     & $7<K_s<11$, $1\simlt J-K_s\simlt1.3$ & {\bf M I-II} \\
I &  21,986 & $0.55$ & $11<K_s\simlt13$, $0.75<J-K_s\simlt1$ & {\bf K-M I-II}, K-M V, M III \\
J &   8,229 & $0$             & $9.5\simlt K_s\simlt11.5$, $1.4\simlt J-K_s<2$ & {\bf C III} \\
K &   2,212 & $0$             & $9\simlt K_s\simlt13$, $2<J-K_s<5$ & {\bf C III} \\
L &   8,940 & $0.01$     & $12.5\simlt K_s<14.75$, $1.2\simlt J-K_s<2.5$ & M late V \\
\hline
\end{tabular}
\vskip 0.1cm
$^{\rm a}$Fraction of Galactic sources estimated from synthetic W92 model

$^{\rm b}$Based on $J-K_s$ color and W92; {\bf LMC} populations in boldface
\end{table*}
\begin{table*}[t]
\caption{LMC stellar populations in the 2MASS CMD \label{table:populations}}
\begin{tabular} {llr}
\hline
Stellar types & Regions$^{\rm a}$ & Typical age \\
\hline
\multicolumn{3}{c}{\large Very Young} \\
\multicolumn{3}{c}{Centrally concentrated, localized to star-forming regions, 
trace spiral structure, weakly trace bar} \\
O3-O6 dwarfs & {\bf A} & $\simlt 10$ Myr \\
Red supergiants, $5-8 M_\odot$ & {\bf H} & $\simlt 50$ Myr \\
Luminous AGB stars, O-rich LPVs, $M\approx5-8 M_\odot$ & {\bf G}, H, F & $\sim 40-100$ Myr \\
Blue and yellow supergiants, LBVs & A, B, C & $<100$ Myr \\
Massive protostars or cocooned OB associations & L & $\simlt 5$ Myr \\
\hline
\multicolumn{3}{c}{\large Young} \\
\multicolumn{3}{c}{Bar-dominated} \\
Luminous E-AGB stars & {\bf G}, H & $200-800$ Myr  \\
Core He-burning giants and supergiants, $M \sim 2-5 M_\odot$, LPVs & {\bf I}, D, C, H & $100-900$ Myr \\
E-AGB, oxygen-rich LPVs & F, E, D & $\simlt 1$ Gyr \\
Carbon stars, C-rich LPVs & J, K & $\simlt 1$ Gyr \\
\hline
\multicolumn{3}{c}{\large Intermediate and Old} \\
\multicolumn{3}{c}{Disk and bar} \\
Low- and intermediate-mass RGB stars & {\bf D}, {\bf E}, L & $1-15$ Gyr \\
O-rich AGB stars, M-S-C stars & {\bf F} & $1-4$ Gyr \\
C-rich TP-AGB stars & {\bf J} & $1-4$ Gyr \\
Dust-enshrouded TP-AGB, carbon stars & K & $1-4$ Gyr \\
\hline
\multicolumn{3}{c}{\large Foreground and Background} \\
\multicolumn{3}{c}{Disk populations and extragalactic component} \\
Disk main-sequence turnoff stars & {\bf B} & $\simlt 7-9$ Gyr \\
Nearby K dwarfs, red clump and red HB stars & {\bf C}, I & $\simlt 9$ Gyr \\
Local F-M dwarfs & {\bf D}, E & varies \\
Background galaxies & {\bf L} & \\
\hline
\end{tabular}
\vskip 0.1cm
$^{\rm a}$Where a population is the dominant contributor to a region, it is
labeled in bold type
\end{table*}
The main points of this preliminary analysis of 2MASS data are the following:
\begin{itemize}
\item The quantity and the quality of 2MASS data allow unprecedented look at 
the entire LMC.  2MASS has produced a rich sample of LMC sources, a few million
stars, with the photometric accuracy of 3-4\%.  $JHK$ 2MASS photometry 
is potentially useful for studying the star-formation 
history of the Cloud.  Cross-correlating 2MASS database with existing
catalogs will provide homogeneous and accurate IR photometry of supergiants
(Sanduleak catalog), Wolf-Rayet stars (Breysacher catalog), Cepheids (OGLE and 
EROS datasets), LBVs and LPVs;
\item The color-color diagram is generally ill-suited to distinguish between
giant (III) and dwarf (V) populations, especially in the color range 
$0.5\le J-K_s\le0.8$.  Nevertheless, the diagram may be useful in identifying 
some candidate LMC objects with infrared excess, such as obscured AGB stars, 
B[e] stars, or LMC protostars.  In addition, the distribution of $J-K_s$ colors 
of a population in a narrow $J-H$ color range is a sensitive reddening test;
\item Major populations can be identified based on the comparison 
of observed CMD features with theoretical positions of known populations.
Isochrone overplotting provides tentative age and metallicity estimates.
We identify a substantial LMC population of AGBs ($\simgt 10^4$ 
sources), and obscured AGBs ($\sim 2000$ sources); 
\item The luminosity function of the LMC giants is determined and tabulated.  We find the
RGB tip at $K_s=12.3\pm0.1$.  Our preliminary analysis of luminosity functions 
in two test fields suggest that luminosity function is the same in the bar and 
the outer regions of the Cloud;
\item Fitting isochrones to the location of the giant branch (including TRGB) 
gives metal abundances consistent between fields.  In particular, we derive 
average metallicity $Z=0.004^{+0.002}_{-0.001}$ for our fields.  Analysis of
deep data gives the same average metallicity.  Our results confirm the absence 
of strong radial metallicity gradient in the field populations of the LMC;
\item The estimates of the distance modulus obtained from our isochrone fits 
to the RGB, are consistent with each other and the most recent results in the 
literature.  The average reddening is marginally different 
between the bar and the outer field, the $E_{B-V}$ for the bar field being
greater.  Distance modulus and reddening estimates from the analysis of deep 
data produces similar values;
\item The ages of dominant RGB populations fall in the range from $3$
to $13$ Gyr, with the average age $\sim 6$ Gyr.  Isochrone fits to the deep
data produce similar age interval, from $3$ to $10$ Gyr, with the average age
$\sim 5$ Gyr.  A more detailed isochrone analysis is required to draw
conclusions about the history of star formation in the LMC  prior to $3-4$ Gyr 
ago;
\item Carbon-rich long-period variables are noted as potential standard
candles.  Due to their significant numbers and narrow luminosity range (which 
may be parametrized through a period-luminosity or luminosity-color relations),
these stars are ideally suited for studying the structure of the LMC along the
line of sight.
\end{itemize}

\section*{Acknowledgements}
It is a pleasure to thank Andrew Cole for his insightful comments and
numerous suggestions which lead to improvement of the paper.  We are
grateful to Michael Skrutskie for his careful reading of the manuscript.
We also
thank Shashi Kanbur for discussions regarding pulsating variables and 
Peter Wood for advice on long-period variables.  This publication makes 
use of data products from the Two Micron All Sky Survey, which is a joint 
project of the University of Massachusetts and the Infrared Processing 
and Analysis Center, funded by the National Aeronautics and Space 
Administration and the National Science Foundation.

\end{document}